\documentclass[lettersize,journal]{IEEEtran}
\pdfoutput=1

\usepackage{cite}
\usepackage{graphicx}
\usepackage[justification=centering]{caption}
\usepackage{autobreak}
\usepackage{setspace}
\usepackage{amsmath}
\usepackage{booktabs}
\usepackage{xcolor}
\usepackage{amssymb}

\ifCLASSINFOpdf
\else
\fi
\begin{document}
%

\title{ \huge  UAV-RIS-Aided Space-Air-Ground Integrated Network: Interference Alignment Design and DoF Analysis
	}

	\allowdisplaybreaks[4]
	\author{Jingfu Li, Gaojie Chen, \textit{Senior Member, IEEE}, Tong Zhang, Wenjiang Feng, Weiheng Jiang, Tony Q. S. Quek,  \textit{Fellow, IEEE} and Rahim Tafazolli, \textit{Senior Member, IEEE}

		\thanks{


		Jingfu Li, Gaojie Chen and Rahim Tafazolli are with 5GIC \& 6GIC, Institute for Communication Systems, University of Surrey, Guildford GU2 7XH, UK (jingfu.li@surrey.ac.uk; gaojie.chen@surrey.ac.uk; r.tafazolli@surrey.ac.uk).
			
		 Tong Zhang is with the Department of Electronic Engineering, College of Information Science and Technology, Jinan University, Guangzhou 510632, China (e-mail: zhangt77@jnu.edu.cn).
	
		Wenjiang Feng, and Weiheng Jiang are with the School of Microelectronics and Communication Engineering, Chongqing University, Chongqing 400044, China (e-mail: fengwj@cqu.edu.cn; whjiang@cqu.edu.cn).

		Tony Q. S. Quek is with the Information Systems Technology Design Pillar, Singapore University of Technology and Design, Singapore 487372 (e-mail: tonyquek@sutd.edu.sg).

	   Corresponding author: Gaojie Chen.
}}	
	\maketitle

\maketitle

	
	\begin{abstract}

In space-air-ground integrated networks (SAGIN), receivers experience diverse interference from both the satellite and terrestrial transmitters. The heterogeneous structure of SAGIN poses challenges for traditional interference management (IM) schemes to effectively mitigate interference. To address this, a novel UAV-RIS-aided IM scheme is proposed for SAGIN, where different types of channel state information (CSI) including no CSI, instantaneous CSI, and delayed CSI, are considered. According to the types of CSI, interference alignment, beamforming, and space-time precoding are designed at the satellite and terrestrial transmitter side, and meanwhile, the UAV-RIS is introduced for cooperating interference elimination process. Additionally, the degrees of freedom (DoF) obtained by the proposed IM scheme are discussed in depth when the number of antennas on the satellite side is insufficient. Simulation results show that the proposed IM scheme improves the system capacity in different CSI scenarios, and the performance is better than the existing IM benchmarks without UAV-RIS.

	\end{abstract}
	
	\begin{IEEEkeywords}
		 Space-air-ground integrated network, UAV-RIS, channel state information, degree of freedom, interference alignment scheme
	\end{IEEEkeywords}

%
\IEEEpeerreviewmaketitle

%
%
%
%

\section{Introduction}

\IEEEPARstart{B}{eyond} the fifth-generation (B5G) and sixth-generation (6G) cellular networks are expected to deliver uninterrupted connectivity and pervasive coverage. Consequently, the space-air-ground integrated network (SAGIN) emerges and plays a critical role in enabling B5G and 6G \cite{ Dang,Shen,9987659}. SAGIN integrating satellites, drone networks, and cellular communications, has garnered considerable attention in recent years, owing to its potential to offer seamless connectivity and pervasive coverage. However, the three-dimensional heterogeneous structure introduces more interference \cite{10159025,5876497}. If the additional interference was not resolved, the integration benefits would not be realized, or the performance was even worse than that of separate networks \cite{7169508}.

For mitigating interference in SAGIN, the classical interference alignment scheme (CIA) \cite{Viveck} as one kind of interference management (IM) scheme was widely employed, particularly in large dimensions and multi-user networks, where $K/2$ sum degrees of freedom (DoF) could be achieved for a $K$-user interference channel (IC). Moreover, the CIA scheme was further extended to fit the asymmetric channel strengths, where a multi-layer IA scheme was proposed in \cite{Jinyuan}. Thereafter, multi-cell and multi-user IA design was found in \cite{Hwang}, and other methods based on the IA were developed, such as opportunistic interference alignment(OIA) \cite{6283995} and interference steering (IS) \cite{8697117}. However, these IA designs \cite{Viveck, Jinyuan, Hwang,6283995,8697117} required current channel state information at the transmitter (CSIT), which was infeasible in satellite communications due to feedback delay.

To make IA schemes available in real scenarios, the issue of CSI latency was considered. In specific, the space-time interference alignment (STIA) scheme was proposed and developed in \cite{Maddah-Ali, STIA, Lee1}. In \cite{Maddah-Ali}, the IA scheme under the assumption of completely delayed CSIT was proposed, which achieved $\frac{K}{1/2+1/3+...+1/K}$ DoF in the $K$-user multiple-input single-output (MISO) broadcast channel. In \cite{STIA}, an STIA was designed where it was shown that using partial delayed and current CSIT achieved the same sum-DoF as that by using completely current CSIT.  Thereafter, the STIA was extended to distributed networks such as the two-user IC and multiple-user X channel in \cite{Lee1}. Notably, a more general STIA scheme was designed in \cite{Tengda} for $M \times N$ user MISO X networks to reduce restrictions on the number of transceiver antennas. In the SAGIN network, since the time experienced by part of nodes to obtain CSI was much greater than others, or the CSI was unavailable, the CSI throughout the network was hybrid, which posed challenges for traditional IA schemes. To address this, the paper proposes a novel IM scheme that takes into account the distinct CSIs of different nodes in SAGIN.

In addition to traditional IM schemes, the reconfigurable intelligent surfaces (RIS) as an emerging technology has shown promise in improving various aspects such as data rates, reliability, and energy efficiency \cite{9507508,10012808,9410435}, as well as enhancing interference management capabilities in terrestrial networks \cite{9681803,9783100,9906810}. Initially, for the scenario with $K$ single-antenna transmitter and receiver pairs, it was demonstrated in \cite{9681803} that an active RIS with sufficient elements could effectively eliminate interference from cascaded channels when there were no direct links among users. The research was then extended to multiuser multiple-input single-output (MISO) networks in \cite{9783100}, where machine-learning methods were used to find the interference zero space of direct and cascaded channels, although the RIS could not be designed directly. Subsequently, the exploration of RIS benefits extended to multiuser MIMO networks with rank-deficient interference channels in \cite{9906810}. In this scenario, the design of each element of the RIS was determined to achieve interference elimination. After that, the focus shifted toward extending the application of RIS to SAGIN. In \cite{9947334}, the active RIS was deployed in SAGIN and the precoding matrix and reflecting coefficient matrix were jointly optimized to minimize the transmit power. However, this work only considered a simple model with one single-antenna primary user and one single-antenna secondary user and did not provide a solution to mitigate interference from the dimensionality of interference space. Thus, the use of RIS for SAGIN, which comprises multiple users with multiple antennas, and achieving interference elimination is a pressing issue.

To tackle this issue, this paper is the first work to design an IA scheme for mitigating interference in SAGIN by assisting a UAV-RIS. Through the joint precoding design of the satellite, terrestrial terminals, and UAV-RIS, the proposed scheme effectively manages interference even when some terminals cannot acquire CSI or the latency of the CSI acquisition is much greater than that of other terminals in SAGIN. In the meantime, the DoF of the whole network is investigated with varied number of transceivers. Overall, the main contributions of our paper are as follows:

{\emph{1)}}
We investigate a UAV-RIS assisted SAGIN, which consists of the satellite, UAV-RIS, and the D2D terminals. In this system, UAV-RIS assists the interference elimination process. In addition, the types of CSIs obtained by different terminals are specifically distinguished as no CSI, instantaneous CSI, and delayed CSI, making the proposed IA scheme more general.

{\emph{2)}} 
We design the first IM scheme that considers the types of CSI on the satellite side. Specifically, when no CSI is acquired by the satellite, the D2D transmitters adopt an interference alignment scheme with the help of UAV-RIS to compress the spatial dimension of the interference. When instantaneous CSI is available on the satellite side, the satellite designs a beamforming precoding matrix to avoid the influence on the D2D receivers, and then the D2D transmitters utilize UAV-RIS to further eliminate the remaining interference. When delayed CSI is obtained by the satellite, the satellite and D2D transmitters leverage spatial and temporal resources to achieve joint precoding, where interference of the whole network is mitigated effectively. In addition to designing the IM scheme, the sum DoF of the whole network is analyzed when the number of antennas on the satellite side is insufficient.

{\emph{3)}}
We investigate the impact of different CSI and network configurations on the proposed scheme, thereby determining the strategy of the scheme under different conditions to ensure optimal performance. Then, simulation results compare the proposed IM scheme with benchmarks under the same condition of CSI and network configurations, illustrating the superiority of the proposed scheme. 

The rest of this paper is structured as follows. In Section II, the SAGIN is introduced along with the CSI latency model and DoF definition. Section III presents three theorems regarding the proposed IM with different CSIs. Subsequently, in Sections IV, V, and VI, each theorem is proven, respectively. The performance of the proposed scheme is then compared to benchmarks under various types of CSIs and network configurations in Section VII. Finally, the paper is concluded in Section VIII.

\section{System Model}
Consider the SAGIN consisting of one satellite $s$, one satellite user $c$, and $K_d$ pairs of device-to-device (D2D), where D2D transmitter $ {k_t}\in  \boldsymbol{D}_t=\{ 1, … ,K_t\} $ matches D2D receiver $ {k_r}\in \boldsymbol{D}_r=\{ 1, … ,K_r\}$, as illustrated in Fig. 1. There is one UAV-RIS $r$ for assisting the D2D communication process. Accordingly, the number of antennas of the satellite, satellite user, D2D transmitter, and D2D receiver are defined as $M_s$, $N_c$, $M_t$, and $N_r$, respectively. The UAV carries an active RIS with $L$ elements.

\begin{figure}[!t]
\centering
\includegraphics[width=3in]{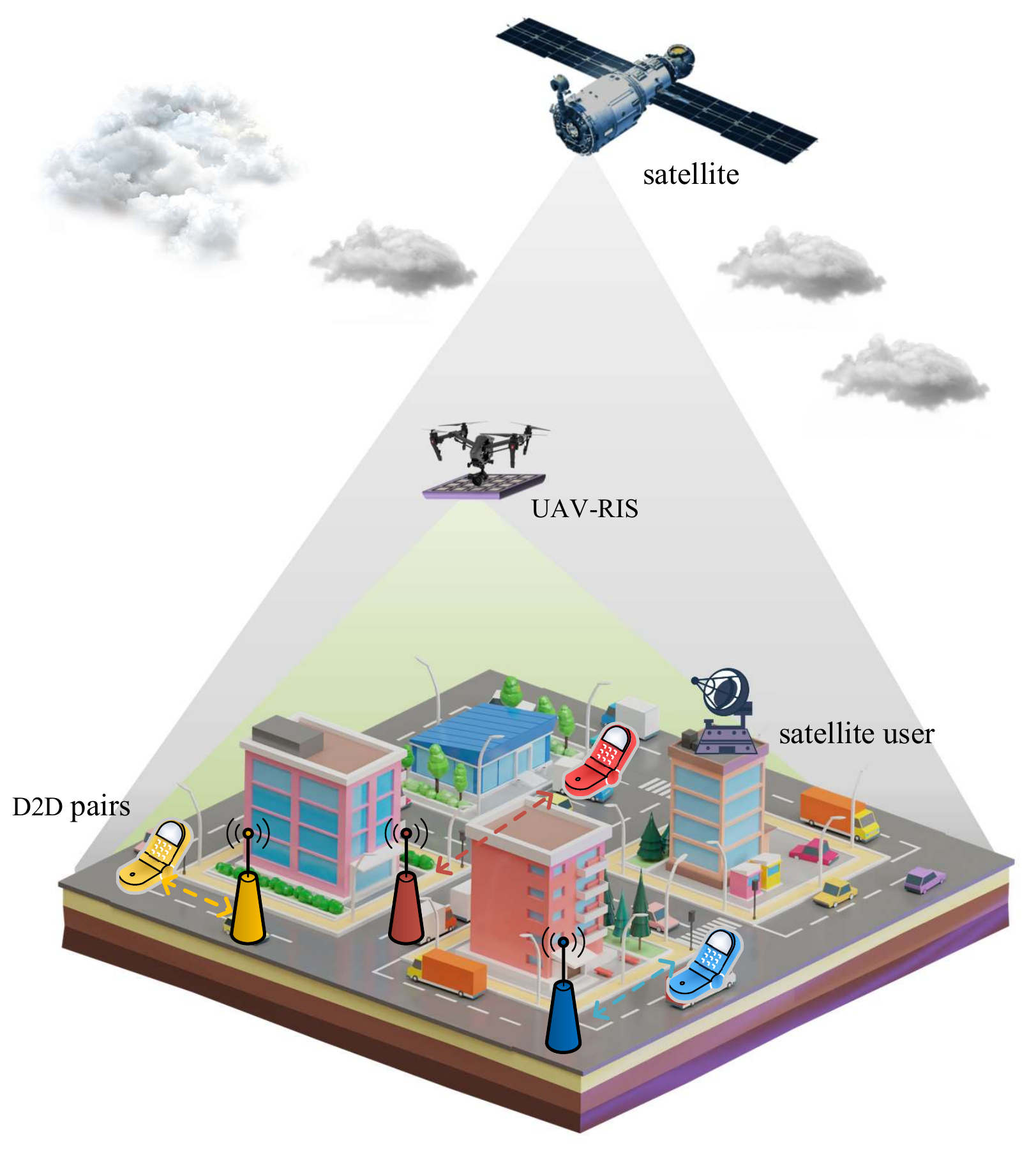}
\caption{SAGIN system model.}
\label{Fig1}
\end{figure}

When the satellite and the $k_t$th D2D transmitter transmit signal ${\boldsymbol{x}_s}\in{\mathbb{C}} {^{M_s \times 1 }}$ and ${\boldsymbol{x}_{k_t}}\in{\mathbb{C}} {^{M_t \times 1 }}$  in the same frequency band, the received signal of the satellite user at the time slot $t$ is
\begin {equation}
\begin{split}
\label {eq1}
\!\!\!\!\!{\boldsymbol{y}_c(t)} &= \sqrt {{P_s}} {\bf{H}}_{s,c}(t){{\bf{W}}_s(t)}{\boldsymbol{x}_s(t)} \\&+ \underbrace {\sum\nolimits_{k_t = 1}^{K_t} {\sqrt {{P_{k_t}}} {\bf{H}}_{k_t,c}(t){{\bf{W}}_{k_t}(t)}{\boldsymbol{x}_{k_t}(t)}} }_{{\rm{interference\, \,from\, \,D2D\, \,pairs}}} \\
&+ \underbrace {\sum \nolimits_{{k_t} = 1}^{K_t} {\sqrt {{P_{k_t}}} {\bf{H}}_{r,c}(t){\bf{\Theta}}(t){\bf{ H}}_{{k_t},r}(t){{\bf{W}}_{k_t}(t)}{\boldsymbol{x}_{k_t}(t)}} }_{{\rm{interference\, \,from\, \,UAV-RIS}}} \\&+ {\boldsymbol{n}_c(t)},
\end{split}
\end {equation}
where ${{\bf{W}}_s(t)}\in{\mathbb{C}} {^{M_s \times  M_s}}$ and ${{\bf{W}}_{k_t}(t)}\in{\mathbb{C}} {^{M_t \times M_t }}$ are the precoding matrices of the satellite and the $k_t$th D2D transmitter. Accordingly, their transmit power are $ {P_s}$ and ${P_{k_t}}$, respectively. ${\bf{H}}_{s,c}(t)\in{\mathbb{C}} {^{N_c \times M_s}}$ denotes the channel matrix from the satellite to satellite user, and it is considered as Shadowed Rician (SR) model, which means that each channel ${{h}}_{{m},{n}}(t)\in{\bf{H}}_{s,c}(t)$ from the $m$th antenna of the satellite to the $n$th antenna of $C$ is subjected to $SR(b, m,\Omega )$, where $SR\left( \cdot , \cdot , \cdot  \right)$ is a random variable with confluent hypergeometric function based probability distribution function. ${\bf{H}}_{{k_t},c}(t)\in{\mathbb{C}} {^{N_c \times M_t }}$, ${\bf{H}}_{r,c}(t)\in{\mathbb{C}} {^{N_c \times L }}$ and ${\bf{H}}_{{k_t},r}(t)\in{\mathbb{C}} {^{L \times M_t }}$ denote the channel matrix from the $k_t$th D2D transmitter to the satellite user, from the UAV-RIS to the satellite user, from the $k_t$th D2D transmitter to the UAV-RIS, respectively, where each channel belonging to these matrices is subjected to Nakagami-m distribution \cite{ref3}. ${{\boldsymbol{n}}_ c}(t) \sim {\cal C}{\cal N}\left( {0,\sigma _c^2} \right)$ is additive white Gaussian noise (AWGN). For the UAV-RIS, its reflecting coefficient matrix ${\bf{\Theta}}\, {{ =  \rm { diag}}}\{ {\tau _1}{{\rm{e}}^{j{\theta _1}}}{\rm{,}} ... {\rm{,}}{\tau _L}{{\rm{e}}^{j{\theta _L}}}\}\in{\mathbb{C}}^ {{L\times L }}$ can be adjusted in the amplitude $\tau $ and the phase $\theta  \in [0,2\pi )$.

Likewise, the received signal vector of the ${k_r}$th D2D receiver at the slot $t$ is
\begin{small}
\begin {equation}
\begin{split}
\label {eq2}
\!\!\!{\boldsymbol{y}_{k_r}}(t) &= \sqrt {{P_{k_t}}} {\bf{H}}_{{k_t},{k_r}}(t){{\bf{W}}_{k_t}}(t){\boldsymbol{x}_{k_t}}(t)\\
 &+ \sqrt {{P_{k_t}}} {\bf{H}}_{r,{k_r}}(t){\bf{\Theta }}(t){\bf{H}}_{{k_t},r}(t){{\bf{W}}_{k_t}}(t){\boldsymbol{x}_{k_t}}(t)\\
 &+ \underbrace {\sqrt {{P_s}} {\bf{H}}_{s,{k_r}}(t){{\bf{W}}_s}(t){\boldsymbol{x}_s}(t)}_{{\rm{interference\,\,from\,\,satellite}}}\\
 &+ \underbrace {\sum\nolimits_{{k_t} = 1\backslash {k_t}}^{K_t} {\sqrt {{P_{k_t}}} {\bf{H}}_{{k_t},{k_r}}(t){{\bf{W}}_{k_t}}(t){\boldsymbol{x}_{k_t}}(t)} }_{{\rm{interference\,\,from\,\,D2D\,\,pairs}}}\\
 &+ \underbrace {\sum\nolimits_{{k_t} = 1\backslash {k_t}}^{K_t} {\sqrt {{P_{k_t}}} {\bf{H}}_{r,{k_r}}(t){\bf{\Theta }}(t){\bf{H}}_{{k_t},r}(t){{\bf{W}}_{k_t}}(t){\boldsymbol{x}_{k_t}}(t)} }_{{\rm{interference\,\,from\,\,UAV-RIS}}}\\
 &+ {\boldsymbol{n}_{k_r}}(t),
\end{split}
\end {equation}
\end{small}
\!\!\!where the channel from the satellite to the $k_r$th D2D receiver,  ${\bf{H}}_{s,{k_r}}(t)$, are subjected to SR distribution. On the other hand, the channels from the $k_t$th D2D transmitter to the $k_r$th D2D receiver, ${\bf{H}}_{{k_t},{k_r}}(t)$, from the UAV-RIS to the $k_r$th D2D receiver, ${\bf{H}}_{r,{k_r}}(t)$, and from the $k_t$th D2D transmitter to the UAV-RIS, ${\bf{H}}_{{k_t},r}(t)$, are subjected to Nakagami-m distribution. 

It can be seen from (\ref{eq1}) and (\ref{eq2}) that both the satellite user and the ${k_r}$th D2D receiver are unavoidably affected by interference from multiple sources. To eliminate the effects of interference, the types of CSIs are distinguished by the CSI latency model \cite{ref4}, and then a novel IM schemes are designed according to different CSI types.

\subsection{CSI Latency Model}
We assume that the CSIs of all terminals are infinite accuracy CSIs \cite{ref4}, which means the quantization error of CSI does not affect the decision of the decoder. Then, we discuss the relationship between CSI and latency. Herein, the CSI latency model is introduced as shown in Fig. 2,
\begin{figure}[!t]
\centering
\includegraphics[width=3in]{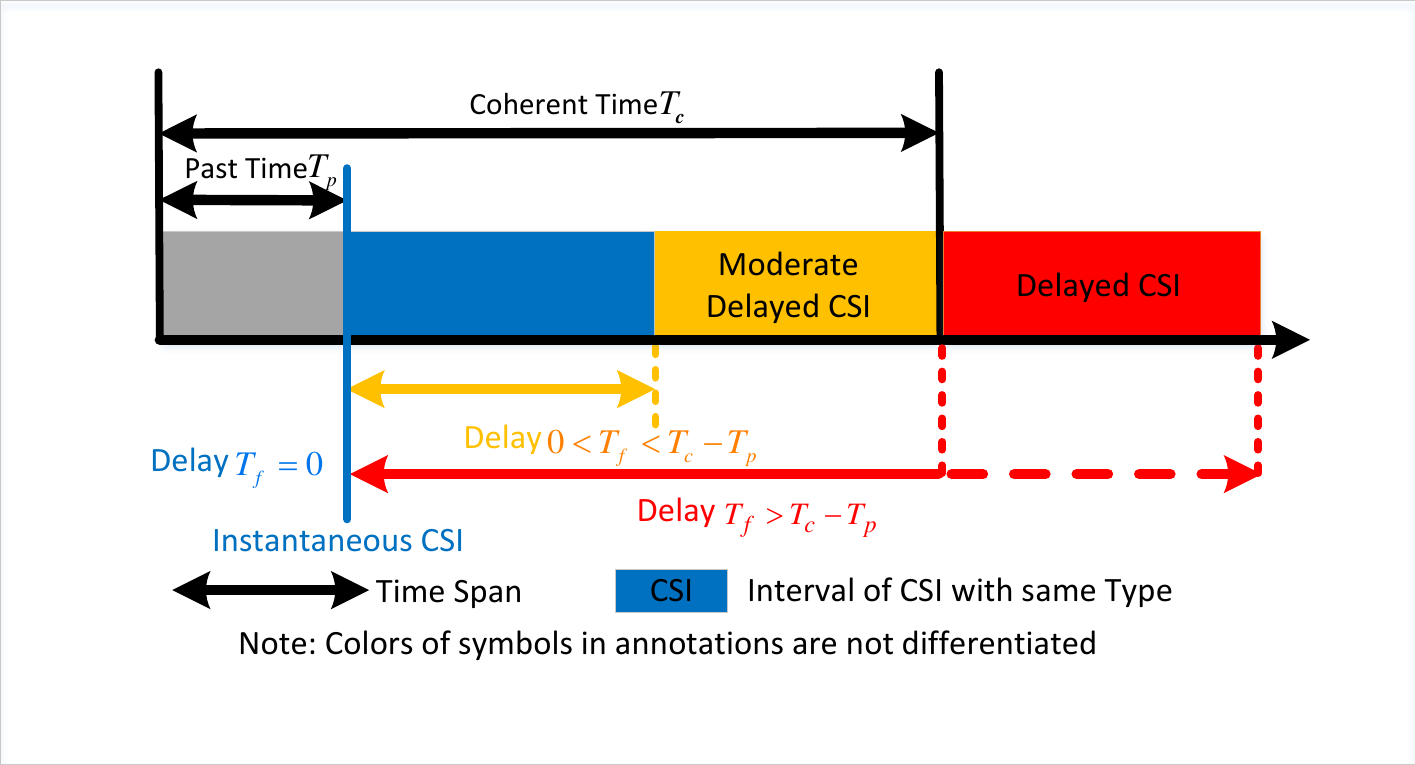}
\caption{\; The illustration of CSI latency model.}
\label{Fig2}
\end{figure}
where ${T_{p}}$, ${T_{f}}$, and ${T_c}$ denote the past time, time delay, and coherence time, respectively. Among them, ${T_{p}}$ is the time elapsed before transmission, and the CSI acquisition process can start at any time within the coherent time slot. The definition of ratio $\lambda$ is 
\begin{equation}
\label{eq3}
\lambda = \frac{{{T_f}}}{{{T_c} - {T_p}}}.
\end{equation}
According to the value of $\lambda$, CSI is determined within three types as follows

\begin{itemize}
\item $\lambda\! =\! 0$ represents instantaneous CSI, where the transmitter receives estimated CSI without delay. This is an upper bound scenario, where transmitters can design precoding matrices to eliminate interference of the current slot. 

\item When $0 \! < \! \lambda \! < \! 1$, two subcases should be considered. The first subcase occurs when the obtained CSI is within the interval of $\left[ {{T_p},{T_p}+{T_f}} \right]$, which is equivalent to the case of instantaneous CSI. The second subcase occurs when the obtained CSI is within the interval of $\left[ {{T_p}+{T_f},{T_c}} \right]$, which is the case of moderately delayed CSI. In the latter case, feedback delay exists but the channel matrix does not change during the remaining time of the coherent time. Thus, the transmitter can design a precoding matrix with moderately delayed CSI to eliminate the interference of the current slot.

\item $\lambda  \ge 1$ represents delayed CSI which is an outdated version. By the time the transmitter obtains CSI, the channel matrix has already changed so that precoding design cannot be made based on delayed CSI. Although the interference of the current slot cannot be completely eliminated with IA technique using delayed CSI, the fixed properties of both the delayed CSI and the channel matrix of past slots can be utilized. In other words, delayed CSI is used for the precoding matrix design of the current slot to eliminate the interference of the past slots.
\end{itemize}

\subsection{Sum Degrees of Freedom}
For the performance of the network, we mainly focus on the system capacity, which is measured by the sum DoF. For the $k$th user, its achievable rate is 
\begin{equation}
\label{eq4}
{R_{k}}{ = }{f_{k}}\log_2 (1{ + }SNR) + o\left( {\log_2 (SNR)} \right),
\end{equation}
where $o(\cdot)$ is the indefinitely small quantity, $SNR$ denotes signal-to-noise ratio. ${f_{k}}$ denotes the DoF of the $k$th user, whose value can be expressed as
\begin{equation}
\label{eq5}
{f_k} = \mathop {\lim }\limits_{SNR \to \infty } \frac{{{R_k}}}{{{{\log }_2}(1+SNR)}},
\end{equation}
which measures the channel capacity of a single cell per user, i.e. the number of data streams that is transmitted reliably by one user per slot. Then, we extend it to the whole network as
\begin{equation}
\label{eq6}
DoF = \sum\limits_{k = 1}^{K} {\mathop {\lim }\limits_{SNR \to \infty } \frac{{{R_k}}}{{{{\log }_2}(1+SNR)}}}. 
\end{equation}
Since the condition $SNR \to \infty$ holds, noise can be ignored in the design of IM scheme.

\section{Main Results: the Achievable DoF of the Considered Network }

\noindent{\bf{Theorem 1.}}
{\emph {For the scenario containing one satellite user and $K_d$ D2D pairs, each of which with $N$ antennas,  $N_c=M_t=N_r=N$. The satellite with $M_s$ antennas obtains no CSI from the satellite to terrestrial terminals. D2D transmitters, D2D receivers, and the satellite user obtain instantaneous CSI from the satellite to themselves, and they also obtain instantaneous CSI from D2D transmitters to all terrestrial terminals. The sum DoF of the whole network is achievable:}}
\begin {equation}
\label {eq7}
DoF =\frac{(K_d + 1)N}{2}.
\end {equation}
{\emph {The specific proof is detailed in Section $\rm{IV}$.}}

\noindent{\bf{Theorem 2.}}
{\emph {Distinguishing from Theorem 1, the satellite obtains the instantaneous CSI from the satellite to terrestrial terminals, and the rest of the assumptions remain unchanged. The sum DoF of the whole network is achievable:}}
\begin {equation}
\label {eq8}
\!\!DoF = \left\{ {\begin{array}{*{20}{c}}
{\frac{(K_d+ 1)N}{2}}&{{M_s} \le \psi }\\
{\left\lceil {\frac{{{M_s}}}{{K_d + 1}}} \right\rceil (K_d + 1)}&{\psi  < {M_s} < (K_d + 1)N}\\
{(K_d+1)N}&{{M_s} \ge (K_d + 1)N},
\end{array}} \right.
\end {equation}
{\emph {where $\psi  = ({K_d} + 1)\left\lfloor {\frac{N}{2}} \right\rfloor$. $\left\lfloor \cdot \right\rfloor$ and $ \left\lceil \cdot \right\rceil$ denote rounding down and rounding up operation, respectively. The theorem is proved in Section $\rm{V}$.}}

\noindent{\bf{Theorem 3.}}
{\emph {Distinguishing from Theorem 1, the satellite obtains the delayed CSI from the satellite to terrestrial terminals, and the rest of the assumptions remain unchanged. The sum DoF of the whole network is achievable:}}
\begin {equation}
\label {eq9}
\left\{ {\begin{array}{*{20}{c}}
{\frac{{{M_s} + (\left\lceil \varphi  \right\rceil  - 1){K_d}N}}{{\left\lceil \varphi  \right\rceil  + 1}}}& 3{K_d}-1  \le  2\varphi  + ({K_d}-3)\left\lceil \varphi  \right\rceil\\
{\frac{(K_d+ 1)N}{2}}&{\rm{otherwise}},
\end{array}} \right.
\end {equation}
{\emph {where $\varphi = \frac{{{M_s}}}{N} $. The proof process is presented in Section $\rm{VI}$.}}

\section{The Proposed Scheme for Achieving Theorem 1}
Restricted by the inability of the satellite to obtain CSI, i.e. no CSI, interference from the satellite to D2D receivers is unavoidable. Therefore, we attempt to design a scheme where part of the interference within the D2D network is aligned to the interference space of the satellite network. The satellite and all D2D receivers jointly design precoding matrices as
\begin{equation}
\label{eq10}
{\begin{array}{*{20}{c}}
{{\bf{H}}_{s,1}{{\bf{W}}_s} = {\bf{H}}_{2,1}{{\bf{W}}_2}},\\
 \vdots \\
{{\bf{H}}_{s,{K_r} - 1}{{\bf{W}}_s} = {\bf{H}}_{{K_t},{{K_r} - 1}}{{\bf{W}}_{K_t}}},\\
{{\bf{H}}_{s,{K_r}}{{\bf{W}}_s} = {\bf{H}}_{1,{K_r}}{{\bf{W}}_1}},
\end{array}}
\end{equation}
where ${{\bf{W}}_s}$ is a unit matrix, implying that the satellite no longer relies on CSI to design precoding matrix. In addition, the singular value decomposition (SVD) method is applied to all aligned matrices, e.g. the pre-coded channel matrix  ${\bf{H}}_{k_t,k_r}{{\bf{W}}_{k_t}}$  between the $k_t$th D2D transmitter to the designated $k_r$th D2D receiver is decomposed into
\begin{equation}
\label{eq11}
{\bf{H}}_{k_t,k_r}{{\bf{W}}_{k_t}} = {\bf{A}}_{k_r}{\bf{\Lambda }}_{k_t,k_r}{\left( {{\bf{B}}_{k_t}} \right)^H},
\end{equation}
where ${\bf{A}}_{k_r} \in{\mathbb{C}} {^{N\times N }}$ and ${{\bf{B}}_{k_t} \in{\mathbb{C}} {^{N \times N }}}$ are unitary matrices, and ${\bf{\Lambda }}_{k_t,k_r}$ is a non-negative rectangular diagonal
matrix. For the $k_t$th D2D transmitter and the $k_r$th D2D receiver, let ${{\bf{\bar W}}_{k_t}} = {{\bf{W}}_{k_t}}{{\bf{B}}_{k_t}}$ and ${{\bf{V}}_{k_r}} = {{\bf{A}}_{k_r}^H}$, respectively, where $(\cdot)^H$ is the conjugate transpose. First, without considering UAV-RIS assistance, the signal received by the $k_r$th D2D receiver is
\begin{equation}
\label{eq12}
\begin{split}
{\boldsymbol{{\hat y}}_{k_r}}&={{\bf{V}}_{k_r}}{\boldsymbol{y}_{k_r}} \\
 & = \sqrt {{P_{k_t}}} {{\bf{V}}_{k_r}}{\bf{H}}_{{k_t},{k_r}}{{{\bf{\bar W}}}_{k_t}}{\boldsymbol{x}_{k_t}} 
\\&+ {\sum\nolimits_{{k_t} = 1\backslash \{{k_t},{k_t}+1\}}^{K_t}  {\sqrt {{P_{k_t}}} {{\bf{V}}_{k_r}}{\bf{H}}_{{k_t},{k_r}}{{{\bf{\bar W}}}_{k_t}}{\boldsymbol{x}_{k_t}}} }\\
&+ \sqrt {{P_{{k_t}+1}}} {{\bf{V}}_{k_r}}{\bf{H}}_{{{k_t}+1},{k_r}}{{{\bf{\bar W}}}_{{k_t}+1}}{\boldsymbol{x}_{{k_t}+1}} \\
 &+ \sqrt {{P_s}} {{\bf{V}}_{k_r}}{\bf{H}}_{s,{k_r}}{{{\bf{ W}}}_s}{\boldsymbol{x}_s}.
\end{split}
\end{equation}
Let $\sqrt {{{P}_s}}{{\boldsymbol{x}}_s'}=\sqrt {{P_{{k_t} + 1}}} {{\bf{B}}_{{k_t}+1}}{\boldsymbol{x}_{{k_t} + 1}} + \sqrt {{P_s}} {\boldsymbol{x}_s}$, we have
\begin{equation}
\label{eq12b}
\begin{split}
{\boldsymbol{{\hat y}}_{k_r}}&= \sqrt {{P_{k_t}}}{{\bf{V}}_{k_r}} {\bf{H}}_{{k_t},{k_r}}{{{\bf{\bar W}}}_{k_t}}{\boldsymbol{x}_{k_t}}\\
&+ \sum\nolimits_{{k_t}= 1\backslash \{ {k_t},{k_t} + 1\} }^{K_t}   {\sqrt {{P_{k_t}}}{{\bf{V}}_{k_r}} {\bf{H}}_{{k_t},{k_r}}{{{\bf{\bar W}}}_{k_t}}{\boldsymbol{x}_{k_t}}} \\
 &+ \sqrt {{{P}_s}} {{\bf{V}}_{k_r}}{\bf{H}}_{s,{k_r}}{{\bf{W}}_s}{{\boldsymbol{x}}_s'}.
\end{split}
\end{equation} 
It can be seen that interference from satellites is integrated into the designated interference space such that the spatial dimension of the interference is compressed. Then, we introduce UAV-RIS to deal with remaining interference. The received signal of the $k_r$th D2D receiver is changed to
\begin{small}
\begin{equation}
\label{eq13}
\begin{split}
{\boldsymbol{{\hat y}}_{k_r}} &= \sqrt {{P_{k_t}}}{{\bf{V}}_{k_r}} {\bf{H}}_{{k_t},{k_r}}{{{\bf{\bar W}}}_{k_t}}{\boldsymbol{x}_{k_t}} \\
 &+ \sqrt {{P_{k_t}}} {{\bf{V}}_{k_r}}{\bf{H}}_{r,{k_r}}{\bf{\Theta H}}_{{k_t},r}{{{\bf{\bar W}}}_{k_t}}{\boldsymbol{x}_{k_t}}\\
&+ \sum\nolimits_{{k_t} = 1\backslash \{ {k_t},{k_t} + 1\} }^{K_t} {\sqrt {{P_{k_t}}} {{\bf{V}}_{k_r}}{\bf{H}}_{{k_t},{k_r}}{{{\bf{\bar W}}}_{k_t}}{\boldsymbol{x}_{k_t}}} \\
&+ \sum\nolimits_{{k_t} = 1\backslash \{ {k_t},{k_t} + 1\} }^{K_t} { \sqrt {{P_{k_t}}} {{\bf{V}}_{k_r}}{\bf{H}}_{r,{k_r}}{\bf{\Theta H}}_{{k_t},r}{{{\bf{\bar W}}}_{k_t}}{\boldsymbol{x}_{k_t}}}\\
 &+ \sqrt {{{P}_s}} {{\bf{V}}_{k_r}}{\bf{H}}_{s,{k_r}}{{\bf{W}}_s}{{\boldsymbol{x}}_s'} \\
 &+ \sqrt {{P_{{k_t} + 1}}}{{\bf{V}}_{k_r}} {\bf{H}}_{r,{k_r}}{\bf{\Theta H}}_{{{k_t} + 1},r}{{{\bf{\bar W}}}_{{k_t} + 1}}{\boldsymbol{x}_{{k_t} + 1}}.
\end{split}
\end{equation}
\end{small}
\!\!\!Let ${\bf{\bar H}}_{{k_t},{k_r}} = {{\bf{A}}_{{k_r}}^H}{\bf{H}}_{{k_t},{k_r}}{{\bf{W}}_{k_t}}{\bf{B}}_{k_t}$, ${\bf{\bar F}}_{{k_t},r} = {\bf{H}}_{{k_t},r}{\bf{B}}_{k_t}$ and ${\bf{\bar G}}_{r,{k_r}} = {{\bf{A}}_{{k_r}}^H}{\bf{H}}_{r,{k_r}}$, we have
\begin{small}
\begin{equation}
\label{eq14}
\begin{split}
\!\!\!{\boldsymbol{{\hat y}}_{k_r}} &= \sqrt {{P_{k_t}}} ({\bf{\bar H}}_{{k_t},{k_r}} + {\bf{\bar G}}_{r,{k_r}}{\bf{\Theta \bar F}}_{{k_t},r}){\boldsymbol{x}_{k_t}}\\
 &+ \sum\nolimits_{{k_t} = 1\backslash \{ {k_t},{k_t} + 1\} }^{K_t} {\sqrt {{P_{k_t}}} ({\bf{\bar H}}_{{k_t},{k_r}} + {\bf{\bar G}}_{r,{k_r}}{\bf{\Theta \bar F}}_{{k_t},r})} {\boldsymbol{x}_{k_t}}\\
 &+ \sqrt {{{P}_s}} {{\bf{A}}_{{k_r}}^H}{\bf{H}}_{s,{k_r}}{{\boldsymbol{x}}_s'}+ \sqrt {{P_{{k_t} + 1}}} {\bf{\bar G}}_{r,{k_r}}{\bf{\Theta \bar F}}_{{{k_t} + 1},r}{\boldsymbol{x}_{{k_t} + 1}}.
\end{split}
\end{equation}
\end{small}
\!\!\!Since the satellite cannot design any precoding matrix while the satellite user can utilize moderately delayed CSI, the zero forcing (ZF) method is taken as
\begin{equation}
\label{eq15}
{{\bf{V}}_c} = {({{\bf{H}}_{s,c}^H}{\bf{H}}_{s,c})^{ - 1}}{{\bf{H}}_{s,c}^H}.
\end{equation}
After ZF processing, the received signal of the satellite user is expressed as
\begin{equation}
\label{eq16}
\begin{split}
{\boldsymbol{{\hat y}}_c} &={{\bf{V}}_c}{\boldsymbol{y}_c} \\
 &= \sqrt {{P_s}} {{\bf{V}}_c}{\bf{H}}_{s,c}{{\bf{W}}_s}{\boldsymbol{x}_s}\\
 &+ \sum\nolimits_{{k_t} = 1}^{K_t} \!\!\!{\sqrt {{P_{k_t}}} {{\bf{V}}_c}({\bf{H}}_{{k_t},c} + {\bf{H}}_{r,c}{\bf{\Theta H}}_{{k_t},r})} {{\bf{W}}_{k_t}}{\boldsymbol{x}_{k_t}}.
\end{split}
\end{equation}
Let ${\bf{\tilde H}}_{s,c} = {{\bf{V}}_c}{\bf{H}}_{s,c}{{\bf{W}}_s}$, ${\bf{\bar F}}_{{k_t},r}= {\bf{H}}_{{k_t},r}{{\bf{W}}_{k_t}}$ and ${\bf{\bar G}}_{r,c} = {{\bf{V}}_c}{\bf{H}}_{r,c}$, we have
\begin{equation}
\label{eq17}
\begin{split}
{\boldsymbol{\hat{y}}_c} &= \sqrt {{P_s}} {\bf{\tilde H}}_{s,c}{\boldsymbol{x}_s} \\
 &+ \sum\nolimits_{{k_t}= 1}^{K_t} {\sqrt {{P_{k_t}}} ({\bf{\tilde H}}_{{k_t},c} + {\bf{\bar G}}_{r,c}{\bf{\Theta \bar F}}_{{k_t},r})} {\boldsymbol{x}_{k_t}}.
\end{split}
\end{equation}
Herein, the ultimate goal of the UAV-RIS is determined, i.e. preserve useful signals while eliminating interference in (\ref{eq14}) and (\ref{eq17}). Accordingly, the following conditions related to UAV-RIS should hold
\begin{equation}
\label{eq18}
\begin{split}
{\bf{\bar H}}_{{k_t},{k_r}} + {\bf{\bar G}}_{r,{k_r}}{\bf{\Theta \bar F}}_{{k_t},r} &= 0,\\
{\bf{\bar G}}_{r,{k_r}}{\bf{\Theta \bar F}}_{{k_t} + 1,r} &= 0,\\
{\bf{\tilde H}}_{{k_t},c} + {\bf{\bar G}}_{r,c}{\bf{\Theta \bar F}}_{{k_t},r} &= 0.
\end{split}
\end{equation}
To satisfy (\ref{eq18}), we adjust the ${\bf{\Theta}}\in {\mathbb{C}}^{L \times L}$ as (\ref{eq19}) which is on the top of next page, 
\newcounter{mytempeqncnt}
\begin{figure*}[!t]
\normalsize
\setcounter{mytempeqncnt}{\value{equation}}
\setcounter{equation}{19}
\begin{equation}
\label{eq19}
\underbrace {\left[ {\begin{array}{*{20}{c}}
{ - {\bf{\bar G}}_{r,1}{\bf{\Theta \bar F}}_{1,r}}&{\bf{0}}& \cdots &{{\bf{\bar H}}_{{{K_t} - 1},1}}&{{\bf{\bar H}}_{{K_t},1}}\\
{{\bf{\bar H}}_{1,2}}&{ - {\bf{\bar G}}_{r,2}{\bf{\Theta \bar F}}_{2,r}}& \cdots &{{\bf{\bar H}}_{{{K_t} - 1},2}}&{{\bf{\bar H}}_{{K_t},2}}\\
\vdots & \vdots & \ddots & \vdots & \vdots \\
{{\bf{\bar H}}_{1,{{K_r} - 1}}}&{{\bf{\bar H}}_{2,{{K_r} - 1}}}& \cdots &{ - {\bf{\bar G}}_{r,{{K_r} - 1}}{\bf{\Theta \bar F}}_{{{K_t} - 1},r}}&{\bf{0}}\\
{\bf{0}}&{{\bf{\bar H}}_{2,{K_r}}}& \cdots &{{\bf{\bar H}}_{{{K_t} - 1},{K_r}}}&{ - {\bf{\bar G}}_{r,{K_r}}{\bf{\Theta \bar F}}_{{K_t},r}}\\
{{\bf{\tilde H}}_{1,c}}&{{\bf{\tilde H}}_{2,c}}& \cdots &{{\bf{\tilde H}}_{{{K_t} - 1},c}}&{{\bf{\tilde H}}_{{K_t},c}}
\end{array}} \right]}_{{{\bf{H}}_r}} = - \underbrace {\left[ {\begin{array}{*{20}{c}}
{{\bf{\bar G}}_{r,1}}\\
\vdots \\
{{\bf{\bar G}}_{r,{K_r}}}\\
{{\bf{\bar G}}_{{r,c}}}
\end{array}} \right]}_{{\bf{\bar G}}}{\bf{\Theta }}{\underbrace {\left[ {\begin{array}{*{20}{c}}
{{\bf{\bar F}}_{1,r}}\\
\vdots \\
{{\bf{\bar F}}_{{K_t},r}}
\end{array}} \right]}_{{\bf{\bar F}}}}^T,
\end{equation}
\hrulefill
\vspace*{4pt}
\end{figure*}
where ${{\bf{H}}_r} \in {\mathbb{C}}{^{({K_d} + 1)N \times {K_d}N}}$, ${\bf{\bar G}}\in {\mathbb{C}}^{({K_d} + 1)N \times L}$ and ${\bf{\bar F}}\in {\mathbb{C}}^{L \times {K_d}N}$ are the equivalence matrices of the whole system. The specific value of each element in the ${\bf{\Theta}}$ is determined by taking the vectorization operation \cite{ref7} as
\begin{equation}
\label{eq20}
\begin{split}
{\rm{vec}}({\bf{\bar H}}_{i,j}) &=  - ({{\bf{\bar F}}_{i,r}^T} \otimes {\bf{\bar G}}_{r,j}){\rm{vec}}({\bf{\Theta }})\\
 & =  - ({{\bf{\bar F}}_{i,r}^T} \otimes {\bf{\bar G}}_{r,j}){\boldsymbol{\beta \alpha }},
\end{split}
\end{equation}
where $i \in \{s,1,...,K_t\}$ is one transmitter and $j \in \{c,1,...,K_r\}$ is one receiver. ${\bf{\bar H}}_{i,j}\in {\mathbb{C}}^{N \times N}$, ${{\bf{\bar F}}_{i,r}^T} \in {\mathbb{C}}^{N \times L}$ and ${\bf{\bar G}}_{r,j} \in {\mathbb{C}}^{N \times L}$ denote submatrix of the equivalence matrices.  $(\cdot)^T$ is the transpose operation and $\otimes $ is the Kronecker product. ${\boldsymbol{\beta }} \in {\mathbb{C}^{{L^2} \times L}}$ is the sparse matrix for $\bf{\Theta}$, whose element $\beta _{q,p}$ of the $p$th column and $q$th row  is given by
\begin{equation}
\label{eq21}
{\beta _{q,p}} = \left\{ {\begin{array}{*{20}{c}}
1\\
0
\end{array}{\rm{  }}} \right.\begin{array}{*{20}{c}}
{{\rm{ }}q = pL - L + p}\\
{\rm{otherwise}}.
\end{array}
\end{equation}
 As for the vector ${\boldsymbol{\alpha }} \in {\mathbb{C}^{L \times 1}}$, it determines the compensation value of each antenna per received terminal, which eliminates the effect of interference. We first set ${{\bf{\bar H}}_r} = ({\bf{F}}^T \otimes {\bf{G}}){\boldsymbol{\beta }}$ such that ${\rm{vec}}({{\bf{H}}_r}) =  - {{\bf{\bar H}}_r}{{\boldsymbol{\alpha }}}$. To make sure that each antenna per received terminal can be regulated by an assigned element of UAV-RIS, the condition must hold
\begin{equation}
\label{eq22}
L \ge  {K_d}({K_d} - 1){N^2} + {K_d}{N^2} \ge {{K_d^2}}{N^2}.
\end{equation}
Under this condition, the Moore–Penrose inverse operation \cite{ref8} is taken as
\begin{equation}
\label{eq23}
{\boldsymbol{\alpha }} =  - {{{\bf{\bar H}}_r}^H}{({{\bf{\bar H}}_r}{{{\bf{\bar H}}_r}^H})^{ - 1}}{\rm{vec}}({{\bf{H}}_r}).
\end{equation}
With the help of UAV-RIS, the channel matrix of the whole system $\bf{\bar{H}}$ is reshaped into (\ref{eq24}) which is on the top of next page.
\newcounter{mytempeqncnt2}
\begin{figure*}[!t]
\normalsize
\setcounter{mytempeqncnt}{\value{equation}}
\setcounter{equation}{23}
\begin{small}
\begin{equation}
\label{eq24}
\begin{split}
&{\bf{\bar H}} = {\bf{H}} + {\bf{G\Theta F}}= {\bf{H}} - {{\bf{H}}_r}\\
&= \left[ {\begin{array}{*{20}{c}}
{{\bf{\bar H}}_{1,1} + {\bf{\bar G}}_{r,1}{\bf{\Theta \bar F}}_{1,r}}&{{\bf{\bar H}}_{2,1} + {\bf{\bar H}}_{s,1}}& \cdots &0&0\\
0&{{\bf{\bar H}}_{2,2} + {\bf{\bar G}}_{r,2}{\bf{\Theta \bar F}}_{2,r}}& \cdots &0&0\\
\vdots & \vdots & \ddots & \vdots & \vdots \\
0&0& \cdots &{{\bf{\bar H}}_{{{K_t} - 1},{{K_r} - 1}} + {\bf{\bar G}}_{r,{{K_r} - 1}}{\bf{\Theta \bar F}}_{{{K_t} - 1},r}}&{{\bf{\bar H}}_{{K_t},{{K_r} - 1}} + {\bf{\bar H}}_{s,{{K_r} - 1}}}\\
{{\bf{\bar H}}_{1,{K_r}} + {\bf{\bar H}}_{s,{K_r}}}&0& \cdots &0&{{\bf{\bar H}}_{{K_t},{K_r}} + {\bf{\bar G}}_{r,{K_r}}{\bf{\Theta \bar F}}_{{K_t},r}}\\
0&0& \cdots &0&0
\end{array}} \right]
\end{split}
\end{equation}
\end{small}
\hrulefill
\vspace*{4pt}
\end{figure*}
Herein, the received signals of the $k_r$th D2D receiver and the satellite user are expressed as
\begin{equation}
\label{eq25}
\begin{split}
{\boldsymbol{{\hat y}}_{k_r}} &= \sqrt {{P_{k_t}}}{{\bf{V}}_{k_r}} {\bf{H}}_{{k_t},{k_r}}{{{\bf{\bar W}}}_{k_t}}{\boldsymbol{x}_{k_t}} \\
 &+\sqrt {{P_{k_t}}}{{\bf{V}}_{k_r}}{\bf{H}}_{r,{k_r}}{\bf{\Theta H}}_{{k_t},r}{{{\bf{\bar W}}}_{k_t}}{\boldsymbol{x}_{k_t}} \\
 &+\sqrt {{{P}_s}} {{\bf{V}}_{k_r}} {\bf{H}}_{s,{k_r}}{{\bf{W}}_s}{{\boldsymbol{x}}_s'}
\end{split}
\end{equation}
and
\begin{equation}
\label{eq26}
{\boldsymbol{{\hat y}}_c} = \sqrt {{P_s}} {{\bf{V}}_c}{\bf{H}}_{s,c}{{\bf{W}}_s}{\boldsymbol{x}_s},
\end{equation}
respectively. To eliminate the remaining interference in (\ref{eq25}), the $k_r$th D2D receiver has to sacrifice half of the received antennas or utilize double resources of time slots. If more time slots are taken, the satellite user has to keep silent to avoid creating additional interference.

In summary, each of the D2D pairs and the satellite user realize ${\frac{N}{2}}$ DoF within one slot. Thus, the sum DoF of the whole system is
\begin{equation}
\label{eq27}
 DoF= \frac{(K_d+ 1)N}{2} .
\end{equation}
Herein, the proof of Theorem $1$ is complete. {\hfill $\blacksquare$}

\section{The Proposed Scheme for Achieving Theorem 2}
In this section, the assumption is made that the satellite has sufficient antennas, ${M_s} \ge ({K_d}+1)N$, and instantaneous CSI or moderately delayed CSI is obtained by the satellite. Given this condition, precoding matrices of the proposed scheme are redesigned and the performance is evaluated with varying ${M_s}$.

Specifically, we first consider the network without UAV-RIS and define channel the matrix from the satellite to the satellite user and D2D receivers as
\begin{equation}
\label{eq28}
{{\bf{H}}_s} = {\left[ {\begin{array}{*{20}{c}}{{\bf{H}}_{s,c}}&{{\bf{H}}_{s,{D_k}}}\end{array}} \right]^T},
\end{equation}
where ${\bf{H}}_{s,c} \in {\mathbb{C}}^{N \times ({K_d}+1)N}$ and ${\bf{H}}_{s,{D_k}} \in {\mathbb{C}}^{{K_d}N\times ({K_d}+1)N}$ denote the channel matrix from the satellite to the satellite user and all D2D receivers ${D_k}$, respectively. To avoid the impact of the satellite network on the D2D network, the interference are aligned to the zero space of ${\bf{H}}_{s,{D_k}}$ by precoding matrix ${\bf{W}}_s$ as
\begin{equation}
\label{eq29}
{\bf{H}}_{s,{D_k}}{{\bf{W}}_s}{\boldsymbol{x}_s} = {\boldsymbol{0}}.
\end{equation}
Since the matrix ${\bf{H}}_{s,{D_k}} \in \mathbb{C}^{{K_d}N\times ({K_d}+1)N}$ is a row full rank matrix, the underlying solution ${{\bf{W}}_{s,c}} \in {\mathbb{C}^{({K_d}+1)N\times N}}$ must exist, thereby the homogeneous equations are designed as
\begin{equation}
\label{eq30}
\begin{array}{c}
{\bf{H}}_{s,1}{{\bf{W}}_{s,c}}{\boldsymbol{x}_{s}^{(1)}} = {\bf{0}},\\
 \vdots \\
{\bf{H}}_{s,{k_r}}{{\bf{W}}_{s,c}}{\boldsymbol{x}_{s}^{(k_t)}} = {\bf{0}},\\
 \vdots \\
{\bf{H}}_{s,{K_r}}{{\bf{W}}_{s,c}}{\boldsymbol{x}_{s}^{(K_t)}} = {\bf{0}},
\end{array}
\end{equation}
where ${{\bf{H}}_{s,{k_r}}}\in {\mathbb{C}^{ N \times ({K_d}+1)N}},\forall {k_r} \in \boldsymbol{D}_r$ is the submatrix of ${\bf{H}}_{s,{D_k}}$, and ${{\boldsymbol{x}}_{s}^{(k_t)}}\in {\mathbb{C}^{N \times 1}},\forall {k_t}\in \boldsymbol{D}_t$ is the subvector of ${{\boldsymbol{x}}_s}$. Let ${\boldsymbol{x}_{s}^{(1)}}=...={\boldsymbol{x}_{s}^{(K_t)}}$ and denote as ${\boldsymbol{x}_{s,c}}$, the received signals of the $k_r$th D2D receiver and the satellite user are
\begin{equation}
\begin{split}
\label{eq31}
	{\boldsymbol{y}_{k_r}} &= \sqrt {{P_{k_t}}} {\bf{H}}_{{k_t},{k_r}}{{\bf{W}}_{k_t}}{\boldsymbol{x}_{k_t}} \\
 &+ \sum\nolimits_{{k_t} = 1\backslash {k_t}}^{K_t} \!\!\! {\sqrt {{P_{k_t}}} {\bf{H}}_{{k_t},{k_r}}{{\bf{W}}_{k_t}}{\boldsymbol{x}_{k_t}}},
\end{split}
\end{equation}
and
\begin{equation}
\label{eq32}
\begin{split}
{\boldsymbol{y}_c} &= \sqrt {{P_s}} {{\bf{H}}_{s,c}}{{\bf{W}}_s}{\boldsymbol{x}_s} \\
 &+ \sum\nolimits_{{k_t} = 1}^{{K_t}} {\sqrt {{P_{k_t}}} {{\bf{H}}_{{k_t},c}}{{\bf{W}}_{k_t}}{\boldsymbol{x}_{k_t}}} \\
 &= \sum\nolimits_{{k_t} = 1}^{{K_t} + 1}{{\sqrt {{P_s}} }}{{\bf{H}}_{s,c}^{({k_t})}} {{\bf{W}}_{s,c}}{\boldsymbol{x}_{s,c}} \\
 &+ \sum\nolimits_{{k_t} = 1}^{{K_t}} {\sqrt {{P_{k_t}}} {{\bf{H}}_{{k_t},c}}{{\bf{W}}_{k_t}}{\boldsymbol{x}_{k_t}}} \\
 &= \sqrt {{P_s}} {{{\bf{\bar H}}}_{s,c}}{{\bf{W}}_{s,c}}{\boldsymbol{x}_{s,c}} \\
 &+ \sum\nolimits_{{k_t} = 1}^{{K_t}} {\sqrt {{P_{k_t}}} {{\bf{H}}_{{k_t},c}}{{\bf{W}}_{k_t}}{\boldsymbol{x}_{k_t}}} ,
\end{split}
\end{equation}
respectively. ${{\bf{H}}_{s,c}^{({k_t})}}\in{\mathbb{C}} {^{N\times N }}$ is the ${k_t}$th subcluster of the satellite to the satellite user, and $ {{{\bf{\bar H}}}_{s,c}} = {{\sum\nolimits_{{k_t} = 1}^{{K_t} + 1} {{\bf{H}}_{s,c}^{({k_t})}} }} $. By utilizing SVD method, ${\bf{H}}_{{k_t},{k_r}}$ and ${\bf{\bar H}}_{s,c}$ are decomposed to
\begin{equation}
\label{eq33}
{\bf{H}}_{{k_t},{k_r}} = {\bf{A}}_{{k_r}}{\bf{\Lambda }}_{{k_t},{k_r}} {{\bf{B}}_{k_t} ^H}
\end{equation}
and
\begin{equation}
\label{eq34}
{\bf{\bar H}}_{s,c} = {\bf{A}}_c{\bf{\Lambda }}_{s,c} {{\bf{B}}_s ^H},
\end{equation}
respectively. ${\bf{A}}_{{k_r}} \in{\mathbb{C}} {^{N\times N }}$, ${{\bf{B}}_{k_t} \in{\mathbb{C}} {^{N \times N }}}$, ${\bf{A}}_c\in{\mathbb{C}} {^{N\times N }}$, and ${{\bf{B}}_s \in{\mathbb{C}} {^{N \times N }}}$ are unitary matrices. ${\bf{\Lambda }}_{{k_t},{k_r}}\in{\mathbb{C}} {^{N \times N }}$ and ${\bf{\Lambda }}_{s,c}\in{\mathbb{C}} {^{N \times N }}$ are non-negative rectangular diagonal matrices. Then, the ${k_t}$th D2D transmitter and the satellite respectively take ${\bf{B}}_{k_t}$ and ${\bf{B}}_s$ as precoding matrix, ${{\bf{W}}_{k_t}} = {\bf{B}}_{k_t}$ and ${{\bf{W}}_s} = {\bf{B}}_s$. Meanwhile, the ${k_r}$th D2D receiver and the satellite user respectively take ${{\bf{A}}_{{k_r}}^H}$ and ${{\bf{A}}_c^H}$ as decoding matrices, ${{\bf{V}}_{k_t}} = {{\bf{A}}_{{k_t}}^H}$ and ${{\bf{V}}_c} = {{\bf{A}}_c^H}$, where $(\cdot)^H$ is the conjugate transpose.

Considering the case that UAV-RIS is adopted to resist interference, the received signal of the $k_r$th D2D receiver is changed to
\begin{small}
\begin{equation}
\label{eq35}
\begin{split}
{\boldsymbol{\hat{y}}_{k_r}} &={{\bf{V}}_{k_r}}{\boldsymbol{y}_{k_r}} \\&= \sqrt {{P_{k_t}}}{{\bf{V}}_{k_r}} {\bf{H}}_{{k_t},{k_r}}{{\bf{W}}_{k_t}}{\boldsymbol{x}_{k_t}}\\
 &+\sqrt {{P_{k_t}}}{{\bf{V}}_{k_r}} {\bf{H}}_{r,{k_r}}{\bf{\Theta H}}_{{k_t},r}{{\bf{W}}_{k_t}}{\boldsymbol{x}_{k_t}}\\
 &+ \sum\nolimits_{{k_t} = 1\backslash {k_t}}^{K_t}{\sqrt {{P_{k_t}}} {{\bf{V}}_{k_r}}{\bf{H}}_{{k_r},{k_t}}{\bf{W}}_{k_t}{\boldsymbol{x}_{k_t}} } \\
 &+ \sum\nolimits_{{k_t} = 1\backslash {k_t}}^{K_t}{ \sqrt {{P_{k_t}}} {{\bf{V}}_{k_r}}{\bf{H}}_{r,{k_r}}{\bf{\Theta H}}_{{k_t},r}{\bf{W}}_{k_t}{\boldsymbol{x}_{k_t}}}. 
\end{split}
\end{equation}
\end{small}
\!\!\!Let ${\bf{\bar H}}_{{k_t},{k_r}} = {{\bf{A}}_{{k_r}}^H}{\bf{H}}_{{k_t},{k_r}}{\bf{B}}_{k_t}$, ${\bf{\bar F}}_{{k_t},r} = {\bf{H}}_{{k_t},r}{\bf{B}}_{k_t}$ and ${\bf{\bar G}}_{r,{k_r}} = {{\bf{A}}_{{k_t}}^H}{\bf{H}}_{r,{k_r}}$, we have
\begin{equation}
\label{eq36}
\begin{split}
{\boldsymbol{\hat{y}}_{k_r}} &= \sqrt {{P_{k_t}}} ({\bf{\bar H}}_{{k_t},{k_r}} + {\bf{\bar G}}_{r,{k_r}}{\bf{\Theta \bar F}}_{{k_t},r}){\boldsymbol{x}_{k_t}}\\
 &+ \sum\nolimits_{{k_t}= 1\backslash {k_t}}^{K_t} {\sqrt {{P_{k_t}}} ({\bf{\bar H}}_{{k_t},{k_r}} + {\bf{\bar G}}_{r,{k_r}}{\bf{\Theta \bar F}}_{{k_t},r})} {\boldsymbol{x}_{k_t}}.
\end{split}
\end{equation}
Similarly, received signal of the satellite user is changed to
\begin{equation}
\label{eq37}
\begin{split}
\!\!\!{\boldsymbol{\hat{y}}_c} &={{\bf{V}}_c}{\boldsymbol{y}_c} \\
 &= \sqrt {{P_s}} {{\bf{V}}_c}{\bf{\bar H}}_{s,c}{{\bf{W}}_{s,c}}{\boldsymbol{x}_{s,c}} \\
 &+ \sum\nolimits_{{k_t}= 1}^{K_t} {\sqrt {{P_{k_t}}} {{\bf{V}}_c}({\bf{H}}_{{k_t},c} + {\bf{H}}_{r,c}{\bf{\Theta H}}_{{k_t},r})} {{\bf{W}}_{k_t}}{\boldsymbol{x}_{k_t}}.
\end{split}
\end{equation}
Let ${\bf{\tilde H}}_{s,c} = {{\bf{V}}_c}{\bf{\bar H}}_{s,c}{{\bf{W}}_s}$, ${{\bf{\tilde H}}_{{k_t},c}}={{\bf{V}}_c}{\bf{H}}_{{k_t},c}{\bf{W}}_{{k_t}}$, ${\bf{\bar F}}_{{k_t},r} = {\bf{H}}_{{k_t},r}{{\bf{W}}_{k_t}}$ and ${\bf{\bar G}}_{r,c} = {{\bf{V}}_c}{\bf{H}}_{r,c}$, we have
\begin{equation}
\label{eq38}
\begin{split}
{\boldsymbol{\hat{y}}_c} &= \sqrt {{P_s}} {\bf{\tilde H}}_{s,c}{\boldsymbol{x}_{s,c}} \\&+ \sum\nolimits_{{k_t} = 1}^{K_t} {\sqrt {{P_{k_t}}} ({\bf{\tilde H}}_{{k_t},c} + {\bf{\bar G}}_{r,c}{\bf{\Theta \bar F}}_{{k_t},r})} {\boldsymbol{x}_{k_t}}.
\end{split}
\end{equation}
Herein, the ultimate goal of the UAV-RIS is to realize the following conditions as
\begin{equation}
\label{eq39}
\begin{split}
{\bf{\bar H}}_{{k_t},{k_r}} + {\bf{\bar G}}_{r,{k_r}}{\bf{\Theta \bar F}}_{{k_t},r} &= 0,\\
{\bf{\tilde H}}_{{k_t}, c} + {\bf{\bar G}}_{r,c}{\bf{\Theta \bar F}}_{{k_t},r} &= 0.
\end{split}
\end{equation}
To satisfy (\ref{eq39}), we adjust the ${\bf{\Theta}}\in {\mathbb{C}}^{L \times L}$ as (\ref{eq40}) which is on the top of the next page, 
\newcounter{mytempeqncnt3}
\begin{figure*}[!t]
\normalsize
\setcounter{mytempeqncnt}{\value{equation}}
\setcounter{equation}{39}
\begin{equation}
\label{eq40}
\underbrace {\left[ {\begin{array}{*{20}{c}}
{ - {\bf{\bar G}}_{r,1}{\bf{\Theta \bar F}}_{1,r}}&{{\bf{\bar H}}_{2,1}}& \cdots &{{\bf{\bar H}}_{{{K_t} - 1},1}}&{{\bf{\bar H}}_{{K_t},1}}\\
{{\bf{\bar H}}_{1,2}}&{ - {\bf{\bar G}}_{r,2}{\bf{\Theta \bar F}}_{2,r}}& \cdots &{{\bf{\bar H}}_{{{K_t} - 1},2}}&{{\bf{\bar H}}_{{K_t},2}}\\
\vdots & \vdots & \ddots & \vdots & \vdots \\
{{\bf{\bar H}}_{1,{{K_r} - 1}}}&{{\bf{\bar H}}_{2,{{K_r} - 1}}}& \cdots &{ - {\bf{\bar G}}_{r,{{K_r} - 1}}{\bf{\Theta \bar F}}_{{{K_t} - 1},r}}&{{\bf{\bar H}}_{{K_t},{{K_r} - 1}}}\\
{{\bf{\bar H}}_{1,{K_r}}}&{{\bf{\bar H}}_{2,{K_r}}}& \cdots &{{\bf{\bar H}}_{{{K_t} - 1},{K_r}}}&{ - {\bf{\bar G}}_{r,{K_r}}{\bf{\Theta \bar F}}_{{K_t},r}}\\
{{\bf{\tilde H}}_{1,c}}&{{\bf{\tilde H}}_{2,c}}& \cdots &{{\bf{\tilde H}}_{{{K_t} - 1},c}}&{{\bf{\tilde H}}_{{K_t},c}}
\end{array}} \right]}_{{{\bf{H}}_r}} = - \underbrace {\left[ {\begin{array}{*{20}{c}}
{{\bf{\bar G}}_{r,1}}\\
\vdots \\
{{\bf{\bar G}}_{r,{K_r}}}\\
{{\bf{\bar G}}_{r,c}}
\end{array}} \right]}_{{\bf{\bar G}}}{\bf{\Theta }}{\underbrace {\left[ {\begin{array}{*{20}{c}}
{{\bf{\bar F}}_{1,r}}\\
\vdots \\
{{\bf{\bar F}}_{{K_t},r}}
\end{array}} \right]}_{{\bf{\bar F}}}}^T
\end{equation}
\hrulefill
\vspace*{4pt}
\end{figure*}
where ${{\bf{H}}_r} \in {\mathbb{C}}{^{({K_d} + 1)N \times {K_d}N}}$, ${\bf{\bar G}}\in {\mathbb{C}}^{({K_d} + 1)N \times L}$ and ${\bf{\bar F}}\in {\mathbb{C}}^{L \times {K_d}N}$. For each reflection of UAV-RIS, the adjustment is consistent with Theorem $2$. With the help of UAV-RIS, the channel matrix of the whole system $\bf{\bar{H}}$ is reshaped into (\ref{eq41}) which is on the top of the next page.
\newcounter{mytempeqncnt4}
\begin{figure*}[!t]
\normalsize
\setcounter{mytempeqncnt}{\value{equation}}
\setcounter{equation}{40}
\begin{equation}
\label{eq41}
\begin{split}
{\bf{\bar H}} &= {\bf{H}} + {\bf{G\Theta F}}= {\bf{H}} - {{\bf{H}}_r}\\
&= \left[ {\begin{array}{*{20}{c}}
{{\bf{\bar H}}_{1,1} + {\bf{\bar G}}_{r,1}{\bf{\Theta \bar F}}_{1,r}}&0& \cdots &0&0\\
0&{{\bf{\bar H}}_{2,2} + {\bf{\bar G}}_{r,2}{\bf{\Theta \bar F}}_{2,r}}& \cdots &0&0\\
\vdots & \vdots & \ddots & \vdots & \vdots \\
0&0& \cdots &{{\bf{\bar H}}_{{{k_t} - 1},{{k_r}- 1}} + {\bf{\bar G}}_{r,{{k_r} - 1}}{\bf{\Theta \bar F}}_{{{k_t} - 1},r}}&0\\
0&0& \cdots &0&{{\bf{\bar H}}_{{k_t},{k_r}} + {\bf{\bar G}}_{r,{k_r}}{\bf{\Theta \bar F}}_{{k_t},r}}\\
0&0& \cdots &0&0
\end{array}} \right]
\end{split}
\end{equation}
\hrulefill
\vspace*{4pt}
\end{figure*}
The received signals of the $k_r$th D2D receiver and the satellite user are given by
\begin{equation}
\label{eq42}
{\boldsymbol{\hat{y}}_{k_r}} = \sqrt {{P_{k_t}}} ({\bf{\bar H}}_{{k_t},{k_r}} + {\bf{\bar G}}_{r,{k_r}}{\bf{\Theta \bar F}}_{{k_t},r}){\boldsymbol{x}_{k_t}}
\end{equation}
and
\begin{equation}
\label{eq43}
{\boldsymbol{\hat{y}}_c} = \sqrt {{P_s}} {\bf{\tilde H}}_{s,c}{\boldsymbol{x}_{s,c}},
\end{equation}
respectively. For (\ref{eq42}) and (\ref{eq43}), it can be seen that the interference is completely eliminated so that both the $k_r$th D2D receiver and the satellite user can achieve decoding. 

In summary, if the condition ${{M_s} \ge ({K_d} + 1)N}$ holds, each of the D2D receivers realizes $N$ DoF and the satellite user realizes $N$ DoF in one slot. Thus, the sum DoF of the whole system is $({K_d} + 1)N$.

\emph{Remark 1 (the case that ${\psi  < {M_s} < ({K_d} + 1)N}$):}
Based on the above conclusion, we gradually decrease the number of $M_s$ and study how the Sum DoF changes. Specifically, if $M_s=({K_d} + 1)N-1$, the first channel submatrix from the satellite to the satellite user,  ${{\bf{H}}_{s,c}^{(1)}}\in{\mathbb{C}} {^{N\times N }}$, is changed to
\begin{equation}
\label{eq44}
{{\bf{H}}_{s,c}^{(1)}} = \left[ {\begin{array}{*{20}{c}}
{h_{{1},{1}}}& \cdots &{h_{{{N - 1}},{1}}}&0\\
 \vdots & \ddots & \vdots & 0\\
{h_{1,{N}}}& \cdots &{h_{{{N - 1}},{N}}}&0
\end{array}} \right],
\end{equation}
where $h_{{m},{n}}, m \in \{1,...,N-1\},n \in \{1,...,N\}$ is a channel from the $m$th antenna to the $n$th antenna. Herein, the received signal from the satellite to the satellite user is given by
\begin{equation}
\label{eq45}
\begin{split}
{\boldsymbol{\hat{y}}_c} &= \sqrt {{P_s}} {{\bf{V}}_c}{\bf{H}}_{s,c}{{\bf{W}}_s}{\boldsymbol{x}_s}\\&= \sqrt {{P_s}} {{\bf{V}}_c}{({{\bf{H}}_{s,c}^{(1)}} + {K_d}\sum\nolimits_{{k_t}= 2}^{{K_t} + 1} {{\bf{H}}_{s,c}^{({k_t})}} )}{{\bf{W}}_{s,c}}{\boldsymbol{x}_{s,c}},
\end{split}
\end{equation}
where the rank of ${\bf{H}}_{s,c}$ does not change, i.e. $ {\rm{rank}}({{\bf{V}}_c}({{\bf{H}}_{s,c}^{(1)}} + {K_d}\sum\nolimits_{{k_t} = 2}^{{K_t} + 1} {{\bf{H}}_{s,c}^{({k_t})}} ))= N$ until ${K_d}+1$ antennas is reduced at the satellite side, where the dimension of each submatrix ${{\bf{H}}_{s,c}^{({k_t})}}$ is reduced by one. Then, if the number of $M_s$ keeps decreasing, the satellite has to sacrifice one DoF.

For the $k_r$th D2D receiver, since the antennas lost by the satellite are gradually decreasing in integer multiples of $N$, the $N$th, $2N$th, $...$, $({K_d} +1)N$th antennas, we denote the transmitted symbol of these antennas as ${\boldsymbol{x}}_{s,c}$. It can be seen that the interference caused by the ${\boldsymbol{x}}_{s,c}$ cannot be eliminated, thereby bringing DoF loss. Since these antennas transmit the same symbol, they can be equated to one-dimensional interference, such that brings one DoF loss for the satellite user but ${K_d}+1$ sum DoF loss for all D2D receivers. After decreasing ${K_d}+1$ antennas of the satellite, each D2D receiver needs to silence one received antenna to compromise the required number of $M_s$ for eliminating interference. At this time, the rank of ${\bf{H}}_{s,{D_k}}$ in (\ref{eq28}) becomes ${({K_d} + 1)(N - 1)}$. Accordingly, its zero-space dimension becomes $N - 1$, which means that each D2D receiver can avoid $N-1$ terms of satellite interference.

Overall, when the number of $M_s$ has decreased ${K_d}+1$ to reach the critical point, and then one additional antenna of the satellite is removed, ${K_d}+1$  sum DoF will be lost at once on top of the IM scheme. In this case, the sum DoF of the whole system is given by
\begin{equation}
\label{eq46}
DoF=({K_d}+1)N - \left\lceil {\frac{{{K_d}N - {M_s}}}{{{K_d} + 1}}} \right\rceil ({K_d} + 1).
\end{equation}
Simplifying (\ref{eq44}), we get
\begin{equation}
\label{eq47}
DoF=\left\lceil {\frac{{{M_s}}}{{{K_d} + 1}}} \right\rceil ({K_d} + 1).
\end{equation}

\emph{Remark 2 (the case that ${{M_s} \le \psi }$):}
For (\ref{eq46}), it shows that the sum DoF keeps decreasing as the number of satellite antennas decreases. There is a critical point where the performance of the IM scheme with moderately delayed CSI does not perform as well as that of the IM scheme with no CSI. Herein, we discuss the critical point regarding $M_s$. Specifically, when the condition 
\begin{equation}
\label{eq48}
\left\lceil {\frac{{{M_s}}}{{{K_d} + 1}}} \right\rceil ({K_d} + 1) \le \frac{{({K_d}+1)N}}{2}
\end{equation}
is satisfied, the IM scheme with no CSI is taken. Otherwise, the IM scheme with moderately delayed CSI is the better choice. Simplifying (\ref{eq48}), we get
\begin{equation}
\label{eq49}
\left\lceil {\frac{{{M_s}}}{{{K_d} + 1}}} \right\rceil  \le \frac{N}{2}.
\end{equation}
If the performance of the IM scheme with no CSI is the same as that of the IM scheme with moderately delayed CSI, it is more preferred due to the cost of CSI acquisition. To ensure that the IM scheme with  moderately delayed CSI is selected only when it can achieve better performance, we transform (\ref{eq49}) to
\begin{equation}
\label{eq50}
{M_s} \le  ({K_d} + 1)\left\lfloor {\frac{N}{2}} \right\rfloor.
\end{equation}
Let $\psi  = ({K_d} + 1)\left\lfloor {\frac{N}{2}} \right\rfloor$. When ${M_s} \le \psi$, the IM scheme with no CSI is adopted, which achieves $\frac{({K_d}+1)N}{2}$ sum DoF. Herein, the proof of Theorem 2 is complete.  {\hfill $\blacksquare$}

\section{The Proposed Scheme for Achieving Theorem 3}

After considering the best-case and worst-case scenarios for the satellite acquisition of the CSI, a more realistic scenario is discussed in this section. Since the satellite works in frequency division duplex (FDD) mode, the CSI acquisition at the satellite experiences three time delays, that is the pilot transmission delay from the satellite to the satellite user and all D2D receivers, feedback delay from the satellite user and all D2D receivers to the satellite \cite{9826890}. If the whole time delay exceeds the coherence time, the satellite obtains the delayed CSI. In this case, the precoding matrices of the proposed scheme are adjusted and its performance is further discussed with the different ${M_s}$.

Generally, we assume that the satellite has adequate antennas,  ${M_s} \ge ({K_d}+1)N$. The proposed IM scheme consists of $3$ phases, where the first phase spans one slot, the $1$st slot; the second phase spans ${K_d}$ slots, from the $2$nd slot to $({K_d}+1)$th slot; and the third phase spans one slot, $({K_d}+2)$th slot. In the slot $1$ of the first phase, the satellite transmits signal ${\boldsymbol{x}_s}[1] \in {{\mathbb{C}}^{_{({K_d}+ 1)N \times 1}}}$ to the satellite user while D2D transmitters keep silent. The received signal of the satellite user and any D2D receiver is given by
\begin{equation}
\label{eq51}
{\boldsymbol{y}_c}[1] = \sqrt {{P_s}} {\bf{H}}_{c,s}[1]{\boldsymbol{x}_s}[1]
\end{equation}
and
\begin{equation}
\label{eq52}
{\boldsymbol{y}_{k_r}}[1] = \sqrt {{P_s}}{\bf{H}}_{s,{k_r}}[1]{\boldsymbol{x}_s}[1],
\end{equation}
respectively. Since the matrix ${\bf{H}}_{s,c}[1] \in {\mathbb{C }}{^{N \times ({K_d} + 1)N}}$ is underdetermined, the satellite user cannot achieve decoding within the first slot. At this time, the delayed CSI of the D2D receivers,  ${\bf{H}}_{s,1}[1], ..., {\bf{H}}_{s,{K_r}}[1]$, is obtained by the satellite and meanwhile the received signal of $k_r$,  ${\boldsymbol{y}_{k_r}}[1] $, is fed back to the D2D transmitter $k_t$.

In any slot of the second phase,  $\forall t\{ 2,...,{K_d} + 1\}$, the signal ${\boldsymbol{x}}_s[t] \in \mathbb{C}^{N\times1}$ transmitted from the satellite to the satellite user is  expressed as
\begin{equation}
\label{eq53}
\begin{split}
{\boldsymbol{x}}_s[t]&=\sqrt {{P_s}}\sum\nolimits_{{k_t} = 1{\backslash (t - 1)}}^{K_t} {{\bf{W}}_{s}}[t]{{\bf{H}}_{s,{k_t}}[1]} {\boldsymbol{x}_s}[1] \\&= \sqrt {{P_s}}{{\bf{W}}_{s}}[t]{{\bf{\bar H}}_s}[t]{\boldsymbol{x}_s}[1],
\end{split}
\end{equation}
where ${{\bf{\bar H}}_s}[t] \in {\mathbb{C}^{N \times ({K_d} + 1)N}}$. The satellite user does not decode the signal in this phase, so its decoding matrix is
\begin{equation}
\label{eq54}
{{\bf{V}}_c}[t] = {\bf{E}},
\end{equation}
where ${\bf{E}}\in {\mathbb{C}^{N \times N}}$ is a unit matrix. Meanwhile, the $(t-1)$th D2D transmitter and each of the remaining D2D transmitters ${k_t} \in {\boldsymbol{D}_t}{\backslash (t - 1)}$  respectively transmit signals
\begin{equation}
\label{eq55a}
{\boldsymbol{x}}_{t-1}[t]=\sqrt {{P_s}}{{\bf{W}}_{t-1}}[t]{\bf{H}}_{s,{t-1}}[1]{\boldsymbol{x}_s}[1]
\end{equation}
and
\begin{equation}
\label{eq55}
{\boldsymbol{x}}_{k_t}[t]=\sqrt {{P_{k_t}}}{{\bf{W}}_{k_t}}[t]{\boldsymbol{x}_{k_t}}[t]
\end{equation}
to their served D2D receivers. For any D2D receiver $k_r$, it decodes the signal with 
\begin{equation}
\label{eq56}
{{\bf{V}}_{k_r}}[t] = {({\bf{H}}_{s,{k_r}}[t])^{ - 1}}.
\end{equation}
Herein, the received signal of $k_r$ is expressed as
\begin{small}
\begin{equation}
\label{eq57}
\begin{split}
{{\boldsymbol{\hat y}}_{k_r}}[t] &= {{\bf{V}}_{k_r}}[t]{\boldsymbol{y}_{k_r}}[t] \\&=\sqrt {{P_s}} {{\bf{V}}_{k_r}}[t] {{\bf{H}}_{s,{k_r}}}[t]{{\bf{W}}_{s}}[t]{{{\bf{\bar H}}}_s}[t]{\boldsymbol{x}_s}[1] \\&+ \sqrt {{P_s}}{{\bf{V}}_{k_r}}[t] {{\bf{H}}_{t - 1,{k_r}}}[t]{{\bf{W}}_{t-1}}[t]{{\bf{H}}_{s,t - 1}}[t]{\boldsymbol{x}_s}[1]\\
 &+ \sqrt {{P_s}}{{\bf{V}}_{k_r}}[t] {{\bf{H}}_{r,{k_r}}}[t]{\bf{\Theta }}[t]{{\bf{H}}_{t - 1,r}}[t]{{\bf{W}}_{t-1}}[t]{{\bf{H}}_{s,t - 1}}[t]{\boldsymbol{x}_s}[1] \\&+ \sqrt {{P_{k_t}}}{{\bf{V}}_{k_r}}[t] {{\bf{H}}_{{k_t},{k_r}}}[t]{{\bf{W}}_{k_t}}[t]{\boldsymbol{x}_{k_t}}[t]\\
 &+ \sqrt {{P_{k_t}}}{{\bf{V}}_{k_r}}[t] {{\bf{H}}_{r,{k_r}}}[t]{\bf{\Theta }}[t]{{\bf{H}}_{{k_t},r}}[t]{{\bf{W}}_{k_t}}[t]{\boldsymbol{x}_{k_t}}[t] \\
&+ \sum\nolimits_{{k_t} = 1\backslash \{ {k_t},t - 1\} }^{{K_t}}\!\!\!\!\!\!\!\!\!\!\!\!\!\!\!  {\sqrt {{P_{k_t}}}{{\bf{V}}_{k_r}}[t] {{\bf{H}}_{{k_t},{k_r}}}[t]{{\bf{W}}_{k_t}}[t]{\boldsymbol{x}_{k_t}}[t]} \\
 &+ \sum\nolimits_{{k_t} = 1\backslash \{ {k_t},t - 1\} }^{{K_t}}\!\!\!\!\!\!\!\!\!\!\!\!\!\!\! {\sqrt {{P_{k_t}}}{{\bf{V}}_{k_r}}[t] {{\bf{H}}_{r,{k_r}}}[t]{\bf{\Theta }}[t]{{\bf{H}}_{{k_t},r}}[t]{{\bf{W}}_{k_t}}[t]{\boldsymbol{x}_{k_t}}[t]} ,
\end{split}
\end{equation}
\end{small}
\!\!\!where precoding matrices ${{\bf{W}}_{t-1}}[t]={{\bf{W}}_{s}}[t]={\bf{E}}$, and meanwhile ${{\bf{W}}_{k_t}}[t]$ is designed as (\ref{eq59}),
\begin{figure*}[!t]
\normalsize
\setcounter{mytempeqncnt4}{\value{equation}}
\setcounter{equation}{58}
\begin{equation}
\label{eq59}
{{\bf{W}}_{k_t}}[t] = \left\{ {\begin{array}{*{20}{c}}
{\bf{E}}\\
{{{({\bf{H}}_{{k_t},{k_r}}[t])}^{ - 1}}{\bf{H}}_{s,{k_r}}[t]{{({\bf{H}}_{s,{k_r}}[t - 1])}^{ - 1}}{\bf{H}}_{{k_t},{k_r}}[t - 1]{{\bf{W}}_{k_t}}[t - 1]}
\end{array}{\rm{   }}\begin{array}{*{20}{c}}
{t = 2}\\
{t \ne 2}
\end{array}} \right. 
\end{equation}
\hrulefill
\vspace*{4pt}
\end{figure*}
which is on the top of the next page. Let ${\bf{\tilde H}}_{s,{k_r}}[t] = {{\bf{V}}_{k_r}}[t]{\bf{H}}_{s,{k_r}}[t]{{\bf{W}}_{s}}[t]$, ${\bf{\tilde H}}_{{t-1},{k_r}}[t] = {{\bf{V}}_{k_r}}[t]{\bf{H}}_{{t-1},{k_r}}[t]{{\bf{W}}_{t-1}}[t]$, ${\bf{\bar F}}_{{t-1},r}[t] = {\bf{H}}_{{t-1},r}[t] {{\bf{W}}_{t-1}}[t]$, ${\bf{\tilde H}}_{{k_t},{k_r}}[t]= {{\bf{V}}_{k_r}}[t]{\bf{H}}_{{k_t},{k_r}}[t]{{\bf{W}}_{k_t}}[t]$, ${\bf{\bar F}}_{K,R}[t] = {\bf{H}}_{{k_t},r}[t]{{\bf{W}}_{k_t}}[t]$, ${\bf{\tilde H}}_{{k_t},{k_r}}[t] = {{\bf{V}}_{k_r}}[t]{\bf{H}}_{{k_t},{k_r}}[t]{{\bf{W}}_{k_t}}[t]$, ${\bf{\bar F}}_{{k_t},r}[t] = {\bf{H}}_{{k_t},r}[t]{{\bf{W}}_{k_t}}[t]$ and ${\bf{\bar G}}_{r,{k_r}}[t] = {{\bf{V}}_{k_r}}[t]{\bf{H}}_{r,{k_r}}[t]$, we have
\begin{small}
\begin{equation}
\label{eq57b}
\begin{split}
{{\boldsymbol{\hat y}}_{k_r}}[t] &=\sqrt {{P_s}} {\bf{\tilde H}}_{s,{k_r}}[t]{{{\bf{\bar H}}}_s}[t]{\boldsymbol{x}_s}[1] \\
 &+ \sqrt {{P_s}}{\bf{\tilde H}}_{{t-1},{k_r}}[t]{{\bf{H}}_{s,t - 1}}[t]{\boldsymbol{x}_s}[1]\\
 &+ \sqrt {{P_s}}{\bf{\bar G}}_{r,{k_r}}[t]{\bf{\Theta }}[t]{\bf{\bar F}}_{{t-1},r}[t]{{\bf{H}}_{s,t - 1}}[t]{\boldsymbol{x}_s}[1] \\
 &+ \sqrt {{P_{k_t}}}{\bf{\tilde H}}_{{k_t},{k_r}}[t]{\boldsymbol{x}_{k_t}}[t]\\
 &+ \sqrt {{P_{k_t}}}{\bf{\bar G}}_{r,{k_r}}[t]{\bf{\Theta }}[t]{\bf{\bar F}}_{{k_t},r}[t]{\boldsymbol{x}_{k_t}}[t] \\
 &+ \sum\nolimits_{{k_t} = 1\backslash \{ {k_t},t - 1\} }^{{K_t}} {\sqrt {{P_{k_t}}}{\bf{\tilde H}}_{{k_t},{k_r}}[t]{\boldsymbol{x}_{k_t}}[t]} \\
 &+ \sum\nolimits_{{k_t} = 1\backslash \{ {k_t},t - 1\} }^{{K_t}}  {\sqrt {{P_{k_t}}}{\bf{\bar G}}_{r,{k_r}}[t]{\bf{\Theta }}[t]{\bf{\bar F}}_{{k_t},r}[t]{\boldsymbol{x}_{k_t}}[t]} ,
\end{split}
\end{equation}
\end{small}
\!\!\!As for the satellite user, its received signal is expressed as
\begin{small}
\begin{equation}
\label{eq58}
\begin{split}
{\boldsymbol{\hat{y}}_c}[t] &={{\bf{V}}_c}[t]{\boldsymbol{{y}}_c}[t] \\
&=\sqrt {{P_s}}{{\bf{V}}_c}[t]{\bf{H}}_{s,c}[t]{{\bf{W}}_s}[t]{{{\bf{\bar H}}}_s}[t]{\boldsymbol{x}_s}[1]\\
&+\sqrt {{P_s}}{{\bf{V}}_c}[t]{\bf{H}}_{{t-1},c}[t]{{\bf{W}}_{t-1}}[t]{\bf{H}}_{s,{t-1}}[1]{\boldsymbol{x}_s}[1]\\
&+ \sqrt {{P_s}} {{\bf{V}}_c}[t]({{\bf{H}}_{r,c}}[t]{\bf{\Theta }}[t]{{\bf{H}}_{t - 1,r}}[t]){{\bf{W}}_{t-1}}[t]{{\bf{H}}_{s,t - 1}}[1]{{\boldsymbol{x}}_s}[1]\\
 &+ \sum\nolimits_{{k_t} = 1{\backslash (t - 1)}}^{K_t}  \!\!\!{\sqrt {{P_{k_t}}} {{\bf{V}}_c}[t]{\bf{H}}_{{k_t},c}[t]} {{\bf{W}}_{k_t}}[t]{\boldsymbol{x}_{k_t}}[t]\\
 &+ \sum\nolimits_{{k_t} = 1{\backslash (t - 1)}}^{K_t} \!\!\!{\sqrt {{P_{k_t}}}{{\bf{V}}_c}[t]({\bf{H}}_{r,c}[t]{\bf{\Theta }}[t]{\bf{H}}_{{k_t},r}[t])} {{\bf{W}}_{k_t}}[t]{\boldsymbol{x}_{k_t}}[t].
\end{split}
\end{equation}
\end{small}
Let ${\bf{\tilde H}}_{s,c}[t] = {{\bf{V}}_c}[t]{\bf{H}}_{s,c}[t]{{\bf{W}}_s}[t]$, ${\bf{\tilde H}}_{{t-1},c}[t] = {{\bf{V}}_c}[t]{\bf{H}}_{{t-1},c}[t]{{\bf{W}}_{t-1}}[t]$, ${\bf{\bar F}}_{{t-1},r}[t] = {{\bf{H}}_{t - 1,r}}[t]{{\bf{W}}_{t-1}}[t]$, ${\bf{\tilde H}}_{{k_t},c}[t] ={{\bf{V}}_c}[t]{\bf{H}}_{{k_t},c}[t]{{\bf{W}}_{k_t}}[t]$, ${\bf{\bar F}}_{{k_t},r}[t] = {\bf{H}}_{{k_t},r}[t]{{\bf{W}}_{k_t}}[t]$ and ${\bf{\bar G}}_{r,c}[t] = {{\bf{V}}_c}[t]{\bf{H}}_{r,c}[t]$, we have
\begin{equation}
\label{eq58b}
\begin{split}
{\boldsymbol{\hat{y}}_c}[t] &=\sqrt {{P_s}}{\bf{\tilde H}}_{s,c}[t]{{{\bf{\bar H}}}_s}[t]{\boldsymbol{x}_s}[1] \\
&+\sqrt {{P_s}}{\bf{{\tilde H}}}_{{t-1},c}[t]{\bf{H}}_{s,{t-1}}[1]{\boldsymbol{x}_s}[1]\\
&+ \sqrt {{P_s}} {{\bf{{\bar G}}}_{r,c}}[t]{\bf{\Theta }}[t]{{\bf{\bar F}}_{t - 1,r}}[t]{{\bf{H}}_{s,t - 1}}[1]{{\boldsymbol{x}}_s}[1] \\
&+ \sum\nolimits_{{k_t} = 1{\backslash (t - 1)}}^{K_t}{\sqrt {P_{k_t}}}{\bf{\tilde H}}_{{k_t},c}[t]{\boldsymbol{x}_{k_t}}[t]\\
 &+ \sum\nolimits_{{k_t} = 1{\backslash (t - 1)}}^{K_t} {\sqrt {{P_{k_t}}}}{\bf{\bar G}}_{r,c}[t]{\bf{\Theta }}[t]{\bf{\bar F}}_{{k_t},r}[t]{\boldsymbol{x}_{k_t}}[t].
\end{split}
\end{equation}
To ensure that the satellite user has access to more information about ${\boldsymbol{x}}_s[1]$, while aligning the interference of any D2D receiver to a determined interference space. The ultimate goal of the UAV-RIS is determined as
\begin{equation}
\label{eq-A}
\begin{split}
{{\bf{\tilde H}}_{t - 1,c}}[t] + {{\bf{{\bar G}}}_{r,c}}[t]{\bf{\Theta }}[t]{{\bf{{\bar F}}}_{t - 1,r}}[t] &= 0,\\
{{\bf{\tilde H}}_{t - 1,{k_r}}}[t]- {{\bf{\tilde H}}_{s,{k_r}}}[t]  + {{\bf{{\bar G}}}_{r,{k_r}}}[t]{\bf{\Theta }}[t]{{\bf{{\bar F}}}_{t - 1,r}}[t] &= 0,\\
{{\bf{\tilde{H}}}_{{k_t},c}}[t] + {\bf{\bar G}}_{r,c}[t]{\bf{\Theta }}[t]{\bf{\bar F}}_{{k_t},r}[t] &= 0,\\
{{\bf{\tilde{H}}}_{{k_t},{k_r}}}[t] + {\bf{\bar G}}_{r,{k_r}}[t]{\bf{\Theta }}[t]{\bf{\bar F}}_{{k_t},r}[t] &= 0.
\end{split}
\end{equation}

To satisfy (\ref{eq-A}), we adjust the ${\bf{\Theta}}[t]\in {\mathbb{C}}^{L \times L}$ as (\ref{eq-B}) which is on the top of next page.
\newcounter{mytempeqncnt5}
\begin{figure*}[!t]
\normalsize
\setcounter{mytempeqncnt}{\value{equation}}
\setcounter{equation}{64}
\begin{equation}
\label{eq-B}
\!\!\!\!\!\!\!\underbrace {\left[ {\begin{array}{*{20}{c}}
{ - {{{\bf{\bar G}}}_{r,1}}[t]{\bf{\Theta }}[t]{{{\bf{\bar F}}}_{1,r}}[t]}& \cdots &{{{{\bf{\tilde H}}}_{t - 1,1}}[t] - {{{\bf{\tilde H}}}_{s,1}}[t]}& \cdots &{{{{\bf{\tilde H}}}_{{K_t},1}}[t]}\\
{{{{\bf{\tilde H}}}_{1,2}}[t]}& \cdots &{{{{\bf{\tilde H}}}_{t - 1,2}}[t] - {{{\bf{\tilde H}}}_{s,2}}[t]}& \cdots &{{{{\bf{\tilde H}}}_{{K_t},2}}[t]}\\
 \vdots & \cdots & \vdots & \cdots & \vdots \\
{{{{\bf{\tilde H}}}_{1,{K_r} - 1}}[t]}& \cdots &{{{{\bf{\tilde H}}}_{t - 1,{K_r} - 1}}[t] - {{{\bf{\tilde H}}}_{s,{K_r} - 1}}[t]}& \cdots &{{{{\bf{\tilde H}}}_{{K_t},{K_r} - 1}}[t]}\\
{{{{\bf{\tilde H}}}_{1,{K_r}}}[t]}& \cdots &{{{{\bf{\tilde H}}}_{t - 1,{K_r}}}[t] - {{{\bf{\tilde H}}}_{s,{K_r}}}[t]}& \cdots &{ - {{{\bf{\bar G}}}_{r,{K_r}}}[t]{\bf{\Theta }}[t]{{{\bf{\bar F}}}_{{K_t},r}}[t]}\\
{{{{\bf{\tilde H}}}_{1,c}}[t]}& \cdots &{{{{\bf{\tilde H}}}_{t - 1,c}}[t]}& \cdots &{{{{\bf{\tilde H}}}_{{K_t},c}}[t]}
\end{array}} \right]}_{{{\bf{H}}_r}[t]} =  - {\bf{\bar G}}[t]{\bf{\Theta }}[t]{\bf{\bar F}}[t]
\end{equation}
\hrulefill
\vspace*{4pt}
\end{figure*}
The process of determining the specific value of each element in ${\bf{\Theta}}[t]$ is consistent with theorems 2 and 3. After eliminating interference with UAV-RIS, the received signal of the satellite user and D2D receiver is changed to
\begin{equation}
\label{eq60}
\begin{split}
{{{\boldsymbol{\hat y}}}_{k_r}}[t] &= \sqrt {{P_s}} {{{\bf{\tilde H}}}_{s,{k_r}}}[t]({{{\bf{\bar H}}}_s}[t] + {{\bf{H}}_{s,t - 1}}[t]){{\boldsymbol{x}}_s}[1]\\&+ \sqrt {{P_{k_t}}} ({{{\bf{\tilde H}}}_{{k_t},{k_r}}}[t] + {{{\bf{\bar G}}}_{r,{k_r}}}[t]{\bf{\Theta }}[t]{{{\bf{\bar F}}}_{{k_t},r}}[t]){{\boldsymbol{x}}_{k_t}}[t]
\end{split}
\end {equation}
and
\begin{equation}
\label{eq61}
{\boldsymbol{\hat{y}}_c}[t] = \sqrt {{P_s}} {\bf{H}}_{s,c}[t]{{\bf{\bar H}}_s}[t]{\boldsymbol{x}_s}[t],
\end {equation}
respectively.

In the third phase, each D2D transmitter transmits signal ${{\bf{W}}_{k_t}}[t]{\boldsymbol{x}_{k_t}}[t - 1]$ while the satellite keeps silent. The received signal of ${k_r}$th D2D receiver is
\begin{equation}
\label{eq63}
{{\boldsymbol{\hat{y}}}_{k_r}}[t] = {{\bf{V}}_{k_r}}[t]{\bf{H}}_{{k_t},{k_r}}[t]{{\bf{W}}_{k_t}}[t]{\boldsymbol{x}_{k_t}}[t - 1],
\end {equation}
where $t={K_d}+2$. Herein, the satellite user jointly decodes the received signals within all previous slots as
\begin{small}
\begin{equation}
\label{eq64}
\begin{split}
{{\boldsymbol{\hat{y}}}_c}&= {\left[ {\begin{array}{*{20}{c}}
{{{\boldsymbol{\hat{y}}}_c}[1]}&{{{\boldsymbol{\hat{y}}}_c}[2]}& ... &{{{\boldsymbol{\hat{y}}}_c}[K_d + 1]}
\end{array}} \right]^T}\\
&= {\underbrace {\left[ {\begin{array}{*{20}{c}}
{{\bf{H}}_{s,c}[1]}&{{\bf{H}}_{s,c}[2]{{{\bf{\bar H}}}_s}[2]} & ...&{{\bf{H}}_{s,c}[K_d + 1]{{{\bf{\bar H}}}_s}[K_d+1]}
\end{array}} \right]}_{{\bf{H}}_{s,c}}}^T \\ & \times{\boldsymbol{x}_s}[1],
\end{split}
\end {equation}
\end{small}
\!\!\!where ${{{\bf{H}}_{s,c}}}$ is a $({K_d}+1)N\times({K_d}+1)N $ full rank matrix. The received signals of any D2D receiver within all previous slots are given by (\ref{eq65}), which is on the top of next page. Besides, both (\ref{eq64}) and (\ref{eq65}) can be decoded by the zero forcing method.
\newcounter{mytempeqncnt7}
\begin{figure*}[!t]
\normalsize
\setcounter{mytempeqncnt}{\value{equation}}
\setcounter{equation}{70}
\begin{equation}
\label{eq65}
\begin{split}
{{\boldsymbol{y}}_{k_r}} &= {\left[ {{{\boldsymbol{y}}_{k_r}}[2] - {{\boldsymbol{y}}_{k_r}}[{k_r} + 1],...,{{\boldsymbol{y}}_{k_r}}[{K_d}] - {{\boldsymbol{y}}_{k_r}}[{k_r} + 1],{{\boldsymbol{y}}_{k_r}}[{K_d} + 2]} \right]^T}\\
 &= \left[ {\begin{array}{*{20}{c}}
{{{{\bf{\tilde H}}}_{{k_t},{k_r}}}[2]}& \cdots &{ - {{{\bf{\tilde H}}}_{{k_t},{k_r}}}[{k_r}+1]}& \cdots &0&0\\
 \vdots & \ddots & \vdots & \ddots & \vdots & \vdots \\
 \vdots & \ddots &0& \ddots & \vdots & \vdots \\
 \vdots & \ddots & \vdots & \ddots &0& \vdots \\
0& \cdots &{ - {{{\bf{\tilde H}}}_{{k_t},{k_r}}}[{k_r}+1]}& \ddots &{{{{\bf{\tilde H}}}_{{k_t},{k_r}}}[{K_d}]}&0\\
0& \cdots &0& \cdots &0&{{{{\bf{\tilde H}}}_{{k_t},{k_r}}}[{K_d} + 1]}
\end{array}} \right]\left[ {\begin{array}{*{20}{c}}
{{{\boldsymbol{x}}_{k_t}}[2]}\\
 \vdots \\
{{{\boldsymbol{x}}_{k_t}}[{k_r}+1]}\\
 \vdots \\
{{{\boldsymbol{x}}_{k_t}}[{K_d}]}\\
{{{\boldsymbol{x}}_{k_t}}[{K_d} + 1]}
\end{array}} \right]
\end{split}
\end{equation}
\hrulefill
\vspace*{4pt}
\end{figure*}

To summarize, each of the ${K_d}$ D2D receivers realizes ${({K_d}-1)}N$ DoF and the $C$ realizes $({K_d}+1)N$ DoF within the $({K_d}+2)$ slot. Thus, the sum DoF is given by
\begin{equation}
\label{eq66}
DoF= \frac{{ {(K_d^2+1)}N}}{{{K_d}+ 2}}.
\end{equation}

\emph{Remark 3 (the relationship between $\frac{M_s}{N} $ and sum DoF):}
It can be seen that the increasing $M_s$ brings a positive gain to sum DoF but there is a critical point that need to be discussed, i.e. the performance of the IM scheme with delayed CSI is inferior to that of the IM scheme with no CSI. We first assume that $M_s=({K_d}+1)N$ and then gradually reduce its number. Specifically, if $M_s=({K_d}+1)N-1$, the desired signal of the satellite user in the first phase is changed to
\begin{equation}
\label{eq67}
{\boldsymbol{x}_s}[1] = {[\begin{array}{*{20}{c}}
{{x_{s,1}}},&{{x_{s,2}}} ,&... &,{{x_{s,({K_d} + 1)N - 1}}}
\end{array}]^T},
\end{equation}
which means that one DoF is sacrificed. Fortunately, it does not affect the dimension of ${{{\bf{\bar H}}}_s}[1]{\boldsymbol{x}_s}[1]$ such that there is no loss on DoF in the second phase. The performance remains stable until $N$ antennas at the satellite are removed. It leads to the consequence that one D2D receiver cannot be served so that the duration of the second phase is reduced by one slot, causing the DoF loss. Therefore, we define $\varphi  = \frac{{{M_s}}}{N}$ to differentiate the number of D2D receivers that can be served. For the satellite user and the $k_r$th D2D receiver, ${\left\lceil \varphi  \right\rceil  + 1}$ slots are taken to realize ${{M_s}}$ DoF and ${(\left\lceil \varphi  \right\rceil  - 1)({K_d}-1)N}$. The sum DoF is given by
\begin{equation}
\label{eq68}
DoF= {\frac{{{M_s} + (\left\lceil \varphi  \right\rceil  - 1)({K_d}-1)N}}{{\left\lceil \varphi  \right\rceil  + 1}}}.
\end{equation}
Similar to the Theorem $2$, there is a critical point where the performance of the IM scheme with delayed CSI does not perform as well as that of the IM scheme with no CSI. Therefore, the condition
\begin{equation}
\label{eq69}
\frac{{({K_d}+1)N}}{2} \le \frac{{{M_s} + (\left\lceil \varphi  \right\rceil  - 1)({K_d}-1)N}}{{\left\lceil \varphi  \right\rceil  + 1}}
\end{equation}
must hold. Otherwise, adopting the IM scheme with no CSI is preferred. Simplifying (\ref{eq69}), we get
\begin{equation}
\label{eq70}
 3{K_d}-1  \le  2\varphi  + ({K_d}-3)\left\lceil \varphi  \right\rceil
,
\end{equation}
Herein, the proof of Theorem $3$ is complete. {\hfill $\blacksquare$}

\section{Numerical Results}

In this section, the paper investigates the DoF gain achieved by the proposed UAV-RIS aided IM scheme in a specific scenario with ${K_d}=6$ and $M_s=({K_d}+1)N$. The impact of varying values of ${K_d}$ and $N$ on the performance of the proposed scheme is also analyzed. Furthermore, the paper discusses the performance of the proposed scheme when the number of satellite antennas $M_s$ is insufficient. To provide a basis for comparison, three benchmark schemes with different types of CSI are introduced:

\begin{itemize}
\item Benchmark with no CSIT: This benchmark, referenced from \cite{ref9}, represents the outer bound of DoF considering various transceiver antenna ratios.  However, when the transceiver antenna ratio equals $1$, the benchmark specifies that the DoF gain is limited. Although other works, e.g. blind interference alignment scheme (BIA), achieve a better DoF, it requires that the transceiver antenna ratios are at some special value points \cite{ref10}, or the channel is broadcast channel \cite{ref11}, both of which are not satisfied in the considered network.

\item Benchmark with instantaneous CSIT: This benchmark, referenced from \cite{ref12}, assumes that the desired signal and interference have different time delays so that each user $k_r$ can obtain up to half of the available time resources without interference. However, the latencies of satellite and D2D communications are different, making the assumptions hard to be satisfied.

\item Benchmark with delayed CSIT: This benchmark, referenced from \cite{ref13}, selectively adopts the retrospective interference alignment (RIA) scheme, precoding scheduling redundancy (PSR) scheme, and TDMA groups (TG) scheme based on specific values of transceiver antenna ratios and the number of received terminals to achieve optimal DoF. 
\end{itemize}
For the aesthetics of the figures, the abbreviations ICSI, DCSI, and NCSI are used to represent instantaneous CSI, delayed CSI, and no CSI, respectively. Similarly, PS and BM are used to abbreviate the proposed scheme and the benchmark, respectively.

\subsection{Sum DoF with UAV-RIS}

\begin{figure}[!t]
\centering
\includegraphics[width=3in]{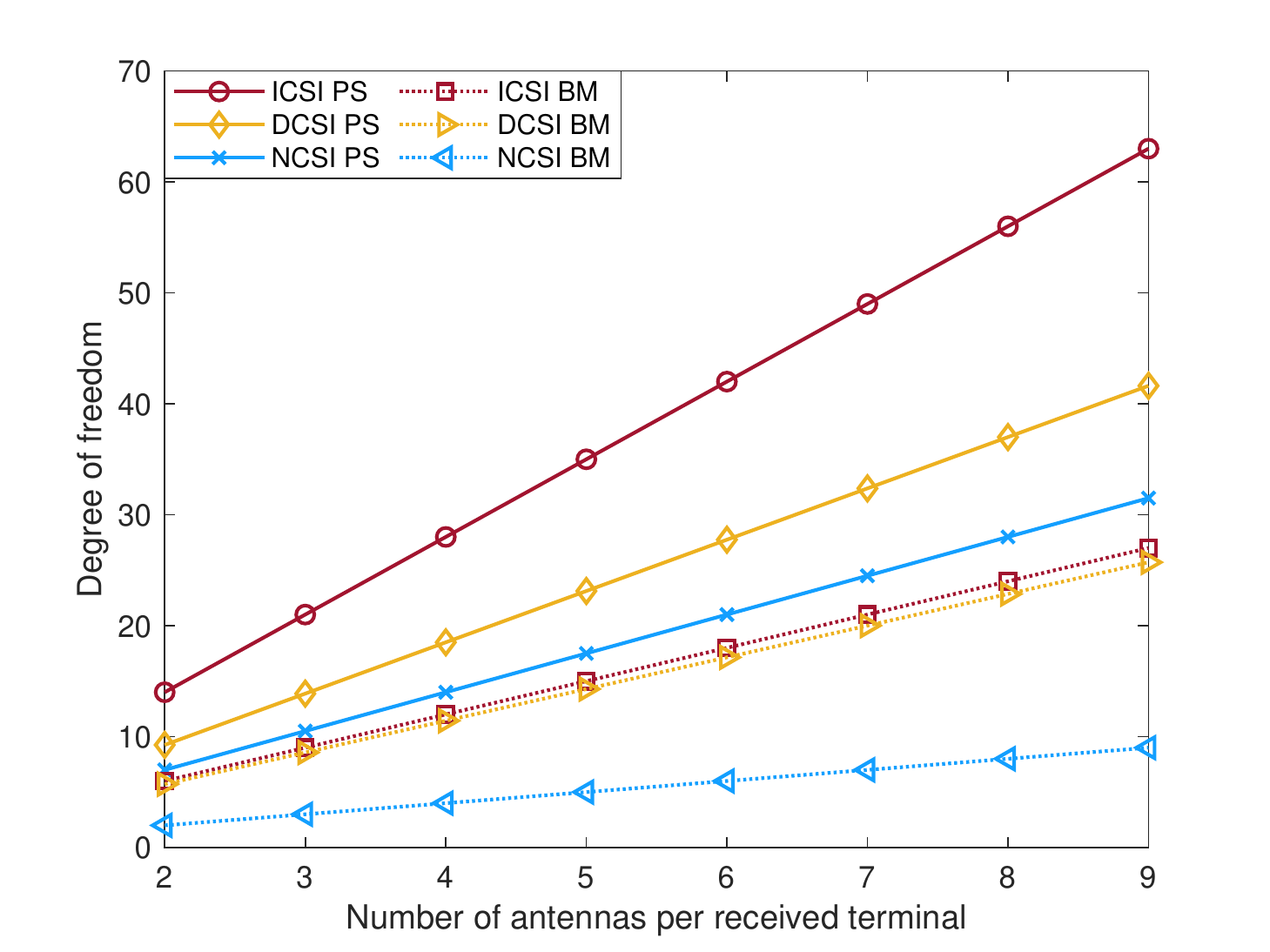}
\caption{The comparison between the proposed schemes and benchmarks in terms of DoF versus $N$, where ${K_d}=6$ and $M_s=({K_d}+1)N$.}
\label{Fig3}
\vspace{-1.0em}
\end{figure}

We set ${K_d}=6$, $M_s=({K_d}+1)N$, and alter the value of $N$ to observe the change in DoF. From Fig. 3, it is observed that all curves show an increase as the number of $N$ is increased, indicating a growing capacity of the system. Since the assumption of sufficient $M_s$ is satisfied, the DoF does not reach its peak and remains stable. Besides, it is seen that both PS and BM achieve the highest DoF gain with ICSI, followed by DCSI, and finally NCSI. This suggests that more ideal CSI results in higher DoF gains, while DCSI can be viewed as a compromise between ICSI and NCSI. Notably, even when PS designs precoding with NCSI, its performance surpasses that of BM with ICSI. This can be attributed to the introduction of UAV-RIS, which assists in communication and allows PS to leverage the favorable characteristics of RIS for effective interference elimination. In conclusion, when sufficient $M_s$ is achieved, PS achieves DoF gains of 133.3\%, 61.9\%, and 250\% compared to BM under the conditions of ICSI, DCSI, and NCSI, respectively.

\subsection{Sum DoF with Adequate $M_s$}

\begin{figure}[!t]
\centering
\includegraphics[width=3in]{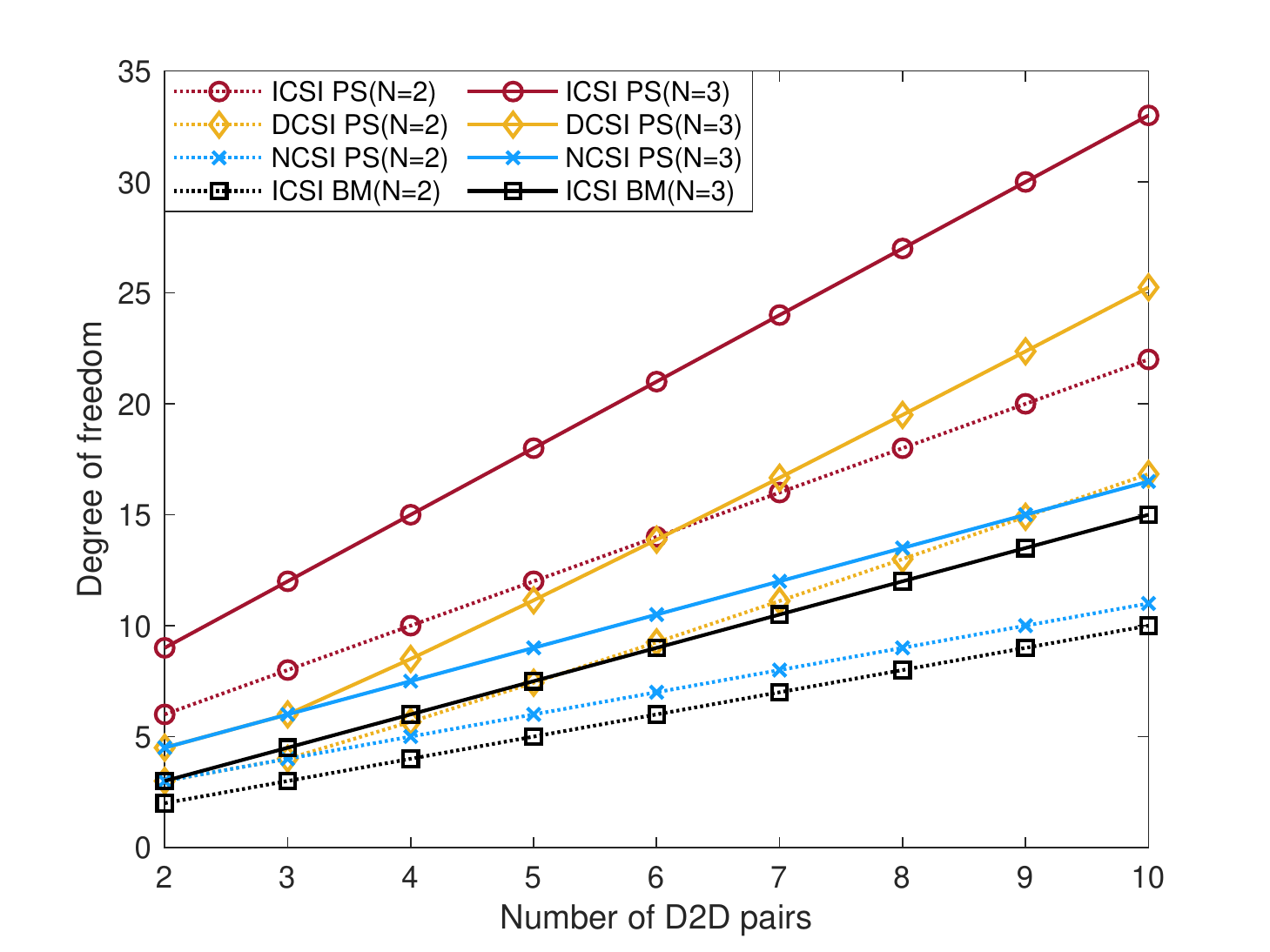}
\caption{The comparison between the proposed schemes and the benchmark with instantaneous CSI in terms of DoF versus $N$ and ${K_d}$, where $M_s=({K_d}+1)N$.}
\label{Fig4}
\end{figure}

Based on the previous analysis, we focus on comparing the ICSI BM with the proposed schemes in the following evaluations. The system configuration remains $M_s=({K_d}+1)N$, and we vary the values of $N$ and ${K_d}$ to study the DoF changes. From Fig. 4, it is evident that the sum DoF increases when there are more D2D pairs or when each D2D pair has more antennas, although the rate of increase differs across schemes. Notably, the change in sum DoF with respect to $N$ is consistent for all schemes, with a $50\%$ DoF gain observed when $N$ ranges from 2 to 3. This can be attributed to the availability of sufficient $M_s$, enabling the design of optimal precoding at the transmitter. On the other hand, when ${K_d}$ varies from 2 to 10, the ICSI PS, DCSI PS, and NCSI PS achieve DoF gains of $266.7\%$, $461.1\%$, and $266.7\%$, respectively. It is worth noting that the DCSI PS exhibits the fastest growth in DoF with respect to ${K_d}$, making it more suitable for multi-user scenarios. For instance, when ${K_d}=10$, the DoF of DCSI PS at $N=2$ even surpasses the DoF of NCSI PS at $N=3$. Overall, regardless of the changes in ${K_d}$ and $N$, the proposed PS scheme with any type of CSI consistently outperforms the ICSI BM in terms of DoF, highlighting the advantageous role of UAV-RIS in assisting communication.

\subsection{Sum DoF with Inadequate $M_s$}

\begin{figure}[!t]
\centering
\includegraphics[width=3in]{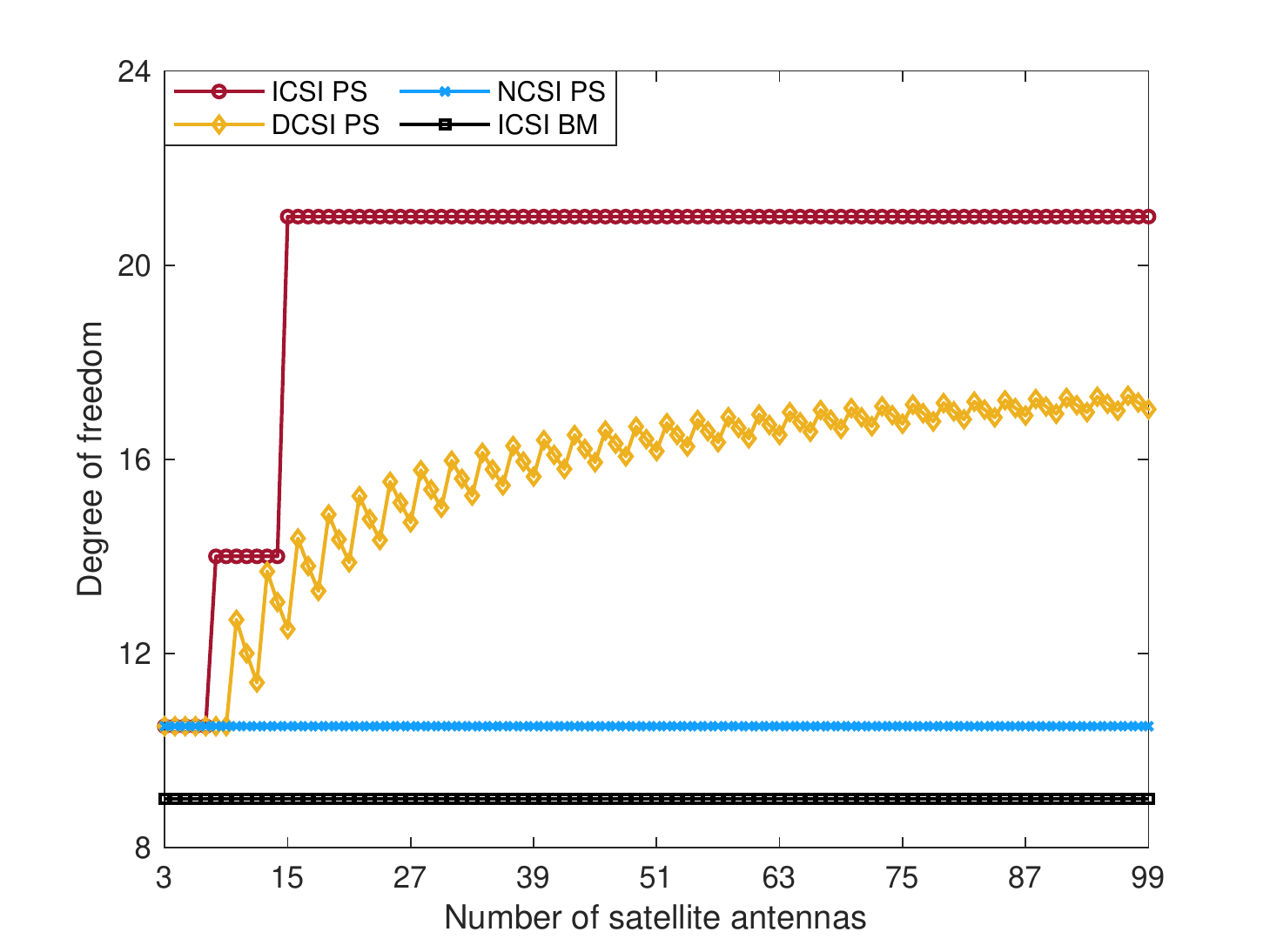}
\caption{The comparison among the proposed schemes in terms of DoF versus $M_s$, where ${K_d}=6$ and $N=3$.}
\label{Fig5}
\end{figure}

The previous conclusions were derived under the assumption of sufficient $M_s$. However, if $M_s$ is insufficient, all PS schemes experience different degrees of DoF loss. To investigate the DoF loss in each PS scheme, we set ${K_d}=6$, $N=3$, and vary the value of $M_s$ from 3 to 99. From Fig. 5, we observe different behaviors for each scheme. For NCSI PS, the DoF remains stable as $M_s$ increases. This is because the satellite cannot design the precoding matrix based on CSI, so a larger $M_s$ does not result in any DoF gain. For DCSI PS, the curve initially remains constant, then suddenly surges and subsequently falls. This behavior repeats in cycles until the curve converges. The constant intervals in DoF arise from the fact that the number of $M_s$ is too small to achieve significant gain, thus the satellite prefers designing the precoding matrix with no CSI. As $M_s$ continues to increase, reaching a critical point adds an additional slot in the second phase of DCSI PS, leading to more efficient spatial utilization. However, adding antennas based on the critical points introduces redundant signals, resulting in new interference towards the received terminals and causing DoF loss. When $M_s$ becomes sufficiently large, the DoF growth tends to approach 0, as it is limited by the outer bound and cannot surpass the inner bound of ICSI PS. For ICSI PS, the curve initially remains constant, then suddenly surges and stabilizes. When the next critical point is reached, the DoF surges again and reaches the inner bound. The reason for the surge is similar to DCSI PS, while the constant DoF intervals occur because the interference caused by the redundant antennas at the satellite side is completely eliminated, preventing DoF loss. In summary, when $M_s$ is at its minimum value, $M_s=N$, all three PS schemes achieve a $16.7\%$ DoF gain compared to the ICSI BM. When $M_s$ reaches 99, whose value is sufficiently large, the PS schemes achieve DoF gains of $133.3\%$, $89.2\%$, and $16.7\%$ based on the ICSI BM for ICSI, DCSI, and NCSI, respectively.

\section{Conclusion}
The study of SAGIN involved the introduction of UAV-RIS to assist in the elimination of interference and the consideration of CSI latency to differentiate the types of CSIs obtained by different terminals. To cope with the interference caused by the complex network structure, a novel IM scheme was proposed. The performance of this scheme was analyzed for the satellite with instantaneous CSI, delayed CSI, and no CSI. Simulation results demonstrated that the proposed IM scheme outperformed the benchmarks in terms of DoF, highlighting the benefits of incorporating UAV-RIS. Additionally, the DoF gain of the proposed scheme was evaluated for different CSIs while maintaining the same system configuration, providing insights into the impact of CSI on performance.

\ifCLASSOPTIONcaptionsoff
  \newpage
\fi

\end{document}